\documentclass{aa}  

\usepackage[colorlinks=true, linkcolor = blue,
            urlcolor  = blue,
            citecolor = blue,
            anchorcolor = blue]{hyperref}
\usepackage{graphicx}
\usepackage{txfonts}
\usepackage{natbib}
\usepackage{enumitem} 
\usepackage{color}
\usepackage{threeparttable}
\usepackage{cancel}
\bibpunct{(}{)}{;}{f}{}{,} 
\usepackage{threeparttable}
\usepackage{xcolor}
\newcommand{\Gaia}{\textit{Gaia}}
\newcommand{\teff}{$T_{\mathrm{eff}}$}
\newcommand{\logg}{log \textit{g}}
\newcommand{\En}{$\mathrm{E}_\mathrm{n}$}

\newcommand{\Lz}{$\mathrm{L}_\mathrm{z}$}

\newcommand{\zmax}{$\mathrm{z}_\mathrm{max}$}

\newcommand{\Jphi}{$\mathrm{J}_\mathrm{\phi}$}

\newcommand{\Jx}{$\mathrm{J}_\mathrm{\bar{x}}$}
\newcommand{\Jy}{$\mathrm{J}_\mathrm{\bar{y}}$}

\newcommand{\Msun}{$\mathrm{M}_\odot$}

\newcommand{\FeH}{$\left[\mathrm{Fe}/\mathrm{H}\right]$}
\newcommand{\BaFe}{$\left[\mathrm{Ba}/\mathrm{Fe}\right]$}
\newcommand{\EuFe}{$\left[\mathrm{Eu}/\mathrm{Fe}\right]$}
\newcommand{\MgFe}{$\left[\mathrm{Mg}/\mathrm{Fe}\right]$}

\newcommand{\BaEu}{$\left[\mathrm{Ba}/\mathrm{Eu}\right]$}

\newcommand{\BPSCS}{BPS CS 29529-0089}
\newcommand{\TYC}{TYC 9219-2422-1}
\newcommand{\HD}{HD 122563}

\begin{document}

  \title{Discovery of a pair of very metal-poor stars enriched in neutron-capture elements:\thanks{Based on observations made with ESO Telescopes at the La Silla Paranal Observatory under programme ID 0108.D-0626(A) and 65.L-0507}}
  \subtitle{The proto-disk r-II star BPS CS 29529-0089 and the Gaia-Sausage-Enceladus r-I star TYC 9219-2422-1}
  
\author{A.~R.~da Silva
          \inst{1}
          \and
          R.~Smiljanic \inst{1}
          }

   \institute{
            Nicolaus Copernicus Astronomical Center, Polish Academy of Sciences, ul. Bartycka 18, 00-716, Warsaw, Poland \\
            \email{arodrigo@camk.edu.pl}
             }

    \date{Received 4 December 2024 / Accepted 26 February 2025}

\authorrunning{da Silva \& Smiljanic}

 
  \abstract
   {R-process enhanced metal-poor stars (\EuFe$\geq+0.3$ and \FeH$\leq-1.0$) are rare objects whose study can provide clues to the astrophysical sites of the rapid neutron capture process.}
   {In this study, we investigate the detailed chemical abundance patterns of two of these anomalous stars, originally identified among stars observed by the GALAH  survey. Our aim is to obtain the detailed chemical abundance pattern of these stars with spectroscopy at higher resolution and signal-to-noise ratio.}
   {We use a calibration of the infrared flux method to determine accurate effective temperatures, and \Gaia~ parallaxes together with broadband photometry and theoretical bolometric corrections to determine surface gravity. Metallicities and chemical abundances are determined with model atmospheres and spectrum synthesis. We also integrate stellar orbits for a complete chemodynamic analysis.}
   {We determine abundances for up to 47 chemical species (44 elements), of which 27 are neutron-capture elements. Corrections because of deviations from the local thermodynamical equilibrium are applied to the metallicities and 12 elements. We find that one of the stars, \BPSCS, is a proto-disk star of the Milky Way of r-II type, with \EuFe=+1.79~dex. The second star, \TYC, is part of the halo and associated with the Gaia-Sausage-Enceladus merger event. It is of r-I type with [Eu/Fe] = +0.54. Abundances of Th are also provided for both stars.}
   {\BPSCS~is the most extreme example of r-process enhanced star known with disk-like kinematics and that is not carbon enhanced. \TYC~is found to be an archetypal Gaia-Sausage-Enceladus star. Their abundances of C, Mg, Ni, Sc, Mn, and Al seem consistent with expectations for stars enriched by a single population III core collapse supernova, despite their relatively high metallicities ([Fe/H] $\sim$ $-$2.4).}


   \maketitle
%
\section{Introduction}

Old, metal-poor stars offer valuable information for understanding the formation and evolution of our Galaxy \citep{Beers2005TheGalaxy}. As long-lived low-mass stars, they are thought to have preserved the chemical properties of the local interstellar medium at the place and time of their formation. One can speculate that the material that formed some of these stars was enriched by only one or maybe a few nucleosynthetic sources \citep{Frebel2015Near-FieldStars}. Therefore, the study of their chemical composition may give unique insight into the first few generations of stars that marked the early history of Galactic chemical enrichment.

Interestingly, some of these old stars have been found to be enriched in r-process elements\citep{Sneden2008Neutron-captureGalaxy.}. The rapid neutron capture process \citep[r process, see e.g.][]{Cowan2021OriginProcess} is a nucleosynthetic mechanism that produces the heaviest elements in the periodic table, along with the slow neutron capture process \citep[s process, e.g.][]{Lugaro2023TheBeyond} and possibly the intermediate neutron capture process \citep[i process, see e.g.][]{Choplin2023TheStars}. The three mechanisms differ with respect to the neutron flux that affects the material and the comparative timescales between successive neutron captures and the beta decay \citep{Arcones2023OriginElements}. Importantly in this context, the astrophysical sources of the r-process elements are still a mystery.

After neutron capture was proposed as the origin of the heavy elements \citep{Burbidge1957SynthesisStars}, several astrophysical sources were suggested as the site where such mechanisms could take place. Among the sites proposed to explain the r process is the merger of degenerate matter objects, such as double systems containing a black hole and a neutron star \citep[BH-NS,][]{Lattimer1974BLACK-HOLE-NEUTRON-STARCOLLISIONS} or two neutron stars \citep[NS-NS,][]{Symbalisty1982NeutronR-Process}. \citet{Li1998TransientMergers} later proposed that such mergers would be accompanied by a transient event, now usually called a kilonova or a macronova. 

In this regard, AT 2017gfo \citep[the optical counterpart of the gravitational wave event GW170817,][]{TheLIGOScientificCollaboration2017GW170817:Inspiral, Coulter2017SwopeSource, Valenti2017The2017gfo/DLT17ck} was the first such event to clearly associate a neutron star merger (NSM) with neutron-capture element production \citep{Tanvir2017TheStars, Pian2017SpectroscopicMerger}. The list of elements with features identified in the AT 2017gfo spectra includes Sr \citep{Watson2019IdentificationStars}, \ion{La}{iii} and \ion{Ce}{iii} \citep{Domoto2022LanthanideKilonovae}, and Te \citep{Hotokezaka2023Tellurium2017gfo}. On the other hand, elements such as Au and Pt were searched but not found \citep{Gillanders2021ConstraintsAT2017gfo}. Currently, there is an active debate about whether lanthanide-poor or lanthanide-rich models better reproduce the spectral evolution of AT 2017gfo \citep{Domoto2021SignaturesSpectra, Vieira2024SpectroscopicKilonova}.  

Other possible sites of the r-process nucleosynthesis include different types of supernovae and hypernovae \citep[see][and references therein]{Cowan2021OriginProcess}. However, there is still no clear observational confirmation of the production of r-process elements at any other astrophysical site. In fact, the topic remains highly debated, and sometimes disparate conclusions are reached. For example, \citet{Siegel2019GW170817R-process} argue that collapsars (high-mass stars that produce a black hole in their collapse, together with a long-duration gamma-ray burst and a hypernova) could supply about 80\% of the r-process elements in the universe. However, the work of \citet{Bartos2019ASystem} indicates that collapsars could possibly contribute only less than 20\% of the r-process elements found in the Solar System. In connection with that, it is worth mentioning that the supernova associated with the long-duration gamma-ray burst GRB 221009A, the brightest observed to date, was found to lack an r-process signature \citep{Blanchard2024JWSTSignature}. 

The current era of large astrometric, photometric, and spectroscopic stellar surveys, such as \textit{Gaia} \citep{GaiaCollaboration2016TheMission}, S-PLUS \citep[Southern Photometric Local Universe Survey,][]{DeOliveira2019TheFilters,Perottoni2024TheRelease}, J-PLUS \citep[Javalambre-Photometric Local Universe Survey,][]{Cenarro2019J-PLUS:Survey}, APOGEE \citep[Apache Point Observatory Galactic Evolution Experiment,][]{Majewski2016TheAPOGEE-2}, \Gaia-ESO \citep{Randich2022TheLegacy} and GALAH \citep[GALactic Archaeology with HERMES,][]{DeSilva2015TheMotivation}, offers the opportunity for the identification and study of larger samples of metal-poor stars \citep[e.g.][]{DaCosta2023SpectroscopicDR3, Placco2023GHOST, Viswanathan2023GaiasSpectra, Bonifacio2024High-speedCandidates, Hou2024VeryDR9, Xylakis-Dornbusch2024MetallicitiesSpectra}. Among these metal-poor stars, it is possible to discover and study objects with unusual r- and s-process abundance patterns, including those that are enriched and/or depleted in such elements \citep[e.g.][]{Francois2007FirstGalaxy, Holmbeck2020TheHalo, Placco2023GHOST, Saraf2023DecodingStars, Sestito2024GHOSTAssembly}. All of these data provide a unique opportunity to obtain a holistic view of all possible sources of the r process.\par

R-process enriched stars (RPE) have traditionally been classified as r-I and r-II \citep{Beers2005TheGalaxy}, where r-I stars are those with moderate enrichment of the r process with $0.3\leq$~\EuFe$~\leq +1.0$~dex and the r-II stars are those with extreme r-process enrichment with \EuFe$~\geq +1.0$~dex. Early studies such as \citet{Sneden1994Ultra--Metal-poor22892-052} and \citet{Hill2002TheCosmochronology} were already able to find a few examples of RPE with quiet high \EuFe~ ratios, $\geq$~+1.6 dex. Nevertheless, such objects are rare. \citet{Ezzeddine2020TheStars} report on a sample of 253 RPE, of which only 7.4\% are r-II. Moreover, recently a new class of super r-process enriched stars, the r-III, was proposed to account for a few of the most extreme objects with \EuFe~$\geq +2.0$~dex \citep{Cain2020TheR-IIIStars, Roederer2024The+2.45}. 

The Stellar Abundances for Galactic Archeology (SAGA) database\footnote{\url{http://sagadatabase.jp/}} \citep{Suda2008StellarStars} lists 26 stars with \EuFe~$\geq +1.7$~dex. Of these, 21 are also carbon-enhanced metal-poor (CEMP, [C/Fe]$\geq+1.0$) stars belonging to the Milky Way. Moreover, \citet{Roederer2024The+2.45} discuss additional four stars that are not in the SAGA database, with \EuFe~$\geq +1.7$~dex: TYC 8100-833-1 and SMSS J145341.38+004046.7 \citep{Ezzeddine2020TheStars}, TYC 7325-920-1 \citep{Cain2020TheR-IIIStars}, and TYC 8444-76-1 \citep{Roederer2024The+2.45}. To our knowledge, the star with the highest nominal \EuFe~ reported to date is BPS CS 31070-0073, a CEMP-r star with \EuFe~= +2.83(25)\footnote{We adopt the SI standard for uncertainties throughout the text. In this notation, the numbers in parenthesis are uncertainties in the last digits of the number, i.e. 3.11(1) is equivalent to 3.11 $\pm$ 0.01 while 3.1(1) is equivalent to 3.1 $\pm$ 0.1.} \citep{Allen2012ElementalStars}.

The vast majority of known RPE are kinematically part of the Milky Way's halo population, with only about 10\% potentially being members of the disk \citep{Gudin2021TheHistory}. Those belonging to the halo include stars formed in situ in the Milky Way as well as accreted stars \citep{Roederer2018KinematicsSatellites, Roederer2018The}. A major source of RPE seems to be the Gaia-Enceladus merger event, a possible origin of about 20\% of the halo RPE \citep{Yuan2020DynamicalHalo, Gudin2021TheHistory, Limberg2021DynamicallySurveys}. The Gaia-Enceladus event, discovered from \textit{Gaia} DR2 data \citep{Belokurov2018Co-formationHalo, Helmi2018TheDisk} is believed to have been the last major merger suffered by the Milky Way \citep[but see][for a different view]{Donlon2024TheYoung}. Gaia-Enceladus completed its merger about 9.6 billion years ago \citep{Giribaldi2023ChronologyPopulations}. On the other hand, \citet{Xie2024Discovery+0.78} recently reported the first RPE with thin disk kinematics which has \EuFe~= +1.32(13) but is not very metal-poor, with \FeH~= $-$0.52. \citet{Johnson2013ChemicalField} found an RPE star in the bulge of the Galaxy. Two other examples of RPE have also been reported in the inner Galaxy and the Bulge, thanks to increased efforts to identify metal-poor stars in these regions \citep{Forsberg2022TheBulge, Mashonkina2023TheStar}. In addition to the Milky Way, other stars with \EuFe~$\geq +1.7$~dex have been found in the dwarf galaxy Reticulum II \citep{Ji2016COMPLETEII}, in the Fornax dwarf galaxy \citep{Reichert2021ExtremeFornax}, in the Large Magellanic Cloud \citep{Reggiani2021TheEnhanced}, and in the Indus stellar stream \citep{Hansen2021Stream}.

In this work, we report a detailed chemical abundance analysis of two new examples of r-process-enhanced very metal-poor stars ([Fe/H] $<$ $-$2). A total of 47 chemical species, including 27 neutron-capture elements, were investigated. These two stars are part of an observational program that we started with the aim of providing detailed chemical abundances to a new set of RPE. As discussed in the next sections, our results indicate that one star (\object{BPS CS 29529-0089}) was probably formed in the proto disk of the Milky Way, while the other (\object{TYC 9219-2422-1}) was probably part of the Gaia-Enceladus system. BPS CS 29529-0089 is the metal-poor star with disk kinematics with highest known \EuFe~that is not a CEMP. This work is organized as follows: we present the selection criteria, observational setups, and analysis methods in Section \ref{sec:Methods}. The results are presented in Section \ref{sec:results} and their implications are discussed in Section \ref{sec:discussion}. Finally, our conclusions are summarized in Section \ref{sec:conclusions}. 

\begin{figure}[!ht]
    \centering
    \includegraphics[width=\linewidth]{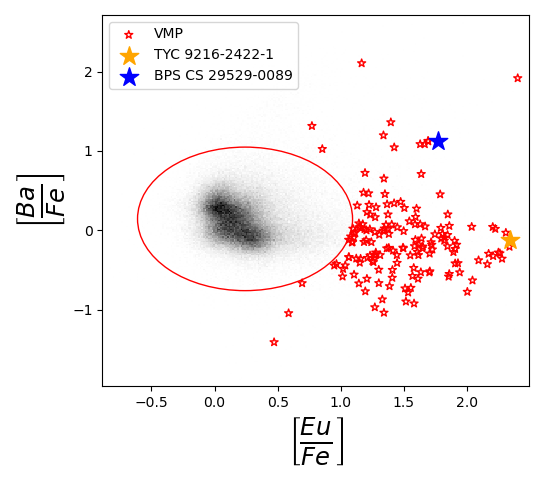}
    \caption{Selection of anomalous stars in GALAH DR3. In gray scale, the 2D histogram of all stars in GALAH DR3 with Ba and Eu abundances. The red ellipsis marks the region differing by 3 $\sigma$ from the mean abundances. The red star symbols show the anomalous stars that also have \FeH~$\leq 2$. The two stars analyzed here are shown in different colors.}
    \label{fig:selection}
\end{figure}

\section{Methods}
\label{sec:Methods}

\subsection{Sample selection and observations}

\begin{table*}[t]
    \centering
    \begin{threeparttable}
    \caption{Observations.}
    \label{tab:logbook}
    \begin{tabular}{ccccccc}
    \hline
    \hline
         Star & Wavelength range & Obs. Date & Res. & S/N & $\rho_{doppler}$ & $\rho_{barycentric}$\\
         Designation & (nm) &  &  &  & (km/s) & (km/s) \\
    \hline
        BPS CS 29529-0089 & 472.6-683.5& 2022-01-18 01:43:09 & 42\,310 & 112 & 91.26(6)  & 84.29(6)\\
        BPS CS 29529-0089 & 328.2-456.3& 2022-01-18 01:43:16 & 40\,970 & 38  & 90.94(3)  & 83.97(3)\\
        BPS CS 29529-0089 & 472.6-683.5& 2022-01-19 03:45:57 & 42\,310 & 113 & 92.52(5)  & 85.58(5)\\
        BPS CS 29529-0089 & 328.2-456.3& 2022-01-19 03:46:01 & 40\,970 & 38  & 92.32(3)  & 85.38(3)\\
        BPS CS 29529-0089 & 472.6-683.5& 2022-02-27 01:27:33 & 42\,310 & 95  & 88.75(6)  & 84.25(6)\\
        BPS CS 29529-0089 & 328.2-456.3& 2022-02-27 01:27:40 & 40\,970 & 33  & 88.83(3)  & 84.33(3)\\
 
        TYC 9219-2422-1 & 472.6-683.5 & 2022-01-18 04:05:23 & 42\,310 & 216  & 209.77(6) & 221.86(6) \\
        TYC 9219-2422-1 & 328.2-456.3 & 2022-01-18 04:05:27 & 40\,970 & 77   & 209.56(4) & 221.65(4) \\
        TYC 9219-2422-1 & 472.6-683.5 & 2022-01-18 04:59:55 & 42\,310 & 190  & 209.61(7) & 221.70(7) \\
        TYC 9219-2422-1 & 328.2-456.3 & 2022-01-18 04:59:59 & 40\,970 & 65   & 209.43(4) & 221.52(4) \\

    \hline
    \end{tabular}
    \begin{tablenotes}
        \item \textit{Notes}. Radial velocities ($\rho$) were determined using {\sf iSpec}~\citep{Blanco-Cuaresma2014DeterminingISpec}.
    \end{tablenotes}
    \end{threeparttable}
\end{table*}

 \begin{table*}[]
    \centering
    \begin{threeparttable}
    \caption{Photometric data.}
    \label{tab:photometricdata}
    \begin{tabular}{cccccc}
    \hline
    \hline
              & BPS CS 29529-0089   & TYC 9219-2422-1 & Source \\
    \Gaia~DR3 & 4678520146255821952 & 5229763049628662784 &  \\
    \hline
        V & 13.105(20) & 11.404(20) & \cite{Zacharias2013TheUCAC4} \\
        B & 13.840(20) & 12.101(20) & \cite{Zacharias2013TheUCAC4} \\
        E($B-V$) & 0.015(1) & 0.117(4) & \cite{Schlafly2010MeasuringSFD}\\
        J & 11.459(27) & 9.819(26) & \cite{Skrutskie2006The2MASS}\\
        H & 10.991(24) & 9.377(23) & \cite{Skrutskie2006The2MASS}\\
        K$_s$ & 10.896(27) & 9.252(23) & \cite{Skrutskie2006The2MASS}\\
        BP & 13.293(3) & 11.592(3) & \cite{GaiaCollaboration2021GaiaProperties}\\
        RP & 12.239(4) & 10.565(4) & \cite{GaiaCollaboration2021GaiaProperties} \\
        BCk & 1.857(6) & 1.591(6) & This work\\
        $\log$(L/L$_\odot$) & 1.9185(9) & 1.1952(4) & This work\\
    \hline
    \end{tabular}
    \begin{tablenotes}
        \item \textit{Notes}. The bolometric correction values for the K$_s$ band (BCk) and luminosity were obtained in an iterative loop following the recipe of \citet{Masana2006EffectivePhotometry}. 
    \end{tablenotes}
    \end{threeparttable}
\end{table*}

The two stars that we discuss here, \BPSCS~ and \TYC, were selected from the GALAH survey. As of data release 3 \citep{Buder2021TheRelease}, the one used in this work, GALAH provides up to 30 elemental abundances for 588\,571 stars, including elemental abundances for two neutron capture elements, Ba and Eu. As the sample of our observational program, we selected very metal-poor ([Fe/H] $\leq$ $-$2) GALAH stars that had values of \EuFe~ and  \BaFe~differing by more than 3 $\sigma$ from the mean of the whole sample (see Fig.~\ref{fig:selection}). For this selection, we did not take into consideration the quality flags of GALAH abundances, as using these flags could exclude outlier stars with uncertain but high Eu and Ba abundances, which are exactly the kind of stars that we are interested in. In total we selected 167 candidates that had Johnson $V$ band magnitudes in the literature. As a pilot study, to test whether we were indeed selecting real examples of RPE, we decided to first observe the two most anomalous and bright stars: TYC 9219-2422-1 and BPS CS 29529-0089.

The two stars were observed with the Ultraviolet and Visual Echelle Spectrograph \citep[UVES,][]{Dekker2000DesignObservatory} at the Very Large Telescope of the European Southern Observatory (ESO). Two spectra were obtained for TYC 9219-2422-1 and three spectra for BPS CS 29529-0089, with resolving power (R) $\sim$ 41\,600 and signal-to-noise ratio (S/N) $\sim$ 50 at 372.4 nm. The wavelength range was divided into two arms, a blue one centered at 390 nm and a red one centered at 580 nm (see details in Tab.~\ref{tab:logbook}). For comparison purposes, we also analyzed spectra of the \Gaia~benchmark star HD~122563 \citep{Jofre2014GaiaMetallicity}. The HD~122563 spectra that we analyzed were taken from the UVES Paranal Observatory Project archive \citep{Bagnulo2003TheDiagram} and observed with spectrograph setups identical to the two stars in our program. Data reduction was performed using the UVES pipeline \citep{Ballester2000ThePipeline}. Using {\sf iSpec} \citep{Blanco-Cuaresma2014DeterminingISpec,Blanco-Cuaresma2019ModernCaveats}, we determined radial velocities ($\rho$) and barycentric corrections, placed them in the rest frame, and combined the multiple spectra of each star.

\subsection{Atmospheric parameters}

The effective temperatures (\teff) were derived iteratively with the Infrared Flux Method (IRFM) calibrations of \citet{Casagrande2021TheSystem} by means of the code {\sf colte}\footnote{\url{https://github.com/casaluca/colte}}, using the photometric bands of the Two Micron All Sky Survey \citep[2MASS,][]{Skrutskie2006The2MASS} and \Gaia~ \citep{Riello2021GaiaValidation}. Values of \teff~determined with such calibrations tend to have a very good accuracy \citep{Giribaldi2021TITANSParallaxes, Giribaldi2023TITANSStars}. The photometric data for \BPSCS~and \TYC~ are given in Tab.~\ref{tab:photometricdata}. 

The values of surface gravity (\logg) were determined with \Gaia~parallaxes \citep{GaiaCollaboration2021GaiaProperties}, corrected by the biases using the tool made available by \citet{Lindegren2021GaiaPosition}, and bolometric correction computed with the method from \citet{Masana2006EffectivePhotometry}. The extinction values were obtained using the Galactic Dust Reddening and Extinction service from the InfraRed Science Archive (IRSA)\footnote{\url{https://irsa.ipac.caltech.edu/applications/DUST/}} and corrected using the recipe from \citet{Beers2002MetalDisk}. 

Metallicities ([Fe/H]) were determined using \ion{Fe}{II} lines that have good values of oscillator strengths( $\log\mathrm{gf}$) given in \citet{Melendez2009BothLines}; see Tab.~\ref{tab:irontwolines} for the list of lines we used. We chose to use \ion{Fe}{ii} lines as they tend to be less sensitive to departures from the local thermodynamical equilibrium (the so-called non-LTE effects) compared to neutral Fe lines. The \ion{Fe}{ii} lines were fitted with synthetic spectra calculated with {\sf TSFitPy}\footnote{\url{https://github.com/TSFitPy-developers/TSFitPy}}, a python wrapper for the Turbospectrum NLTE\footnote{\url{https://github.com/bertrandplez/Turbospectrum_NLTE}} code \citep{Gerber2023Non-LTETurbospectrum,Plez2012Turbospectrum:Synthesis}. For the calculations, we assumed a fixed microturbulence velocity ($\nu_{mic}$) of 1.5 km s$^{-1}$, used MARCS model atmospheres \citep{Gustafsson2008AProperties}, and adopted the solar abundances from \citet{Asplund2009TheSun} modified by \citet{Magg2022ObservationalSun}. The atmospheric parameters determined for our two sample stars are given in Tab.\ \ref{tab:atmosphericparameters}. For HD~122563, we adopted the accurate values of \teff~and \logg~ determined by \citet{Giribaldi2023TITANSStars} with a revised metallicity value\footnote{Giribaldi, R. 2024, private communication}.

Uncertainties in \teff~ are provided by {\sf colte} and come from Monte Carlo simulations using the uncertainties of the photometric bands. For the \logg~uncertainties, we applied a Monte Carlo to the photometric bands, \teff, parallax, and \FeH~over their uncertainties (the standard deviation in the case of metallicity) and added the systematic uncertainty in the bolometric correction presented in \citet{Masana2006EffectivePhotometry}. We adopted a fixed uncertainty for $\nu_{mic}$ of 0.1 km s$^{-1}$. The fixed values adopted here for $\nu_{mic}$ and its uncertainty are consistent with what was found, for example, by \citet{Limberg2024ExtendingWukong/LMS-1} for metal-poor stars of similar surface gravity. For \FeH, we rerun {\sf TSFitPy} to determine the effect of changing the other atmospheric parameters (\teff, \logg~ and $\nu_{mic}$) by their uncertainties.

\begin{table}
    \small
    \centering
    \begin{threeparttable}
    \caption{List of \ion{Fe}{II} lines used in this work.}
    \label{tab:irontwolines}
    \begin{tabular}{cccccc}
    \hline
    \hline
        line    & $\mathrm{\log{gf}}$    & \multicolumn{2}{c}{\BPSCS}   & \multicolumn{2}{c}{\TYC} \\
        (\AA)     &               &       \multicolumn{2}{c}{[Fe/H]}        &      \multicolumn{2}{c}{[Fe/H]}     \\
            &   & LTE & NLTE & LTE & NLTE \\
    \hline
        4178.86 & -2.51 & -2.52 & -2.35 & --     & --    \\
        4233.17 & -1.97 & -2.57 & -2.46 &  -2.32 & -2.23 \\
        4416.83 & -2.65 & -2.63 & -2.35 &  -2.57 & -2.21 \\
        4491.40 & -2.71 & -2.51 & -2.42 &  -2.33 & -2.19 \\
        4508.29 & -2.44 & -2.57 & -2.48 &  -2.44 & -2.43 \\
        4923.93 & -1.26 & -2.41 & -2.53 &  -2.36 & -2.45 \\
        5018.44 & -1.10 & -2.43 & -2.49 &  -2.35 & -2.26 \\
        5197.58 & -2.22 & -2.50 & -2.38 &  -2.34 & -2.21 \\
        5234.62 & -2.18 & -2.43 & -2.63 &  -2.36 & -2.40 \\
        5264.81 & -3.13 & -2.36 & -2.55 &  --    & --    \\
        5414.07 & -3.58 & -2.46 & -2.81 &  -2.08 & -2.53 \\
        5425.26 & -3.22 & -2.47 & -2.32 &  --    & --    \\
        6432.68 & -3.57 & -2.21 & -2.29 &  -2.29 & -2.12 \\
        6456.38 & -2.05 & --    & --    &  -1.93 & -1.85 \\
        6516.08 & -3.31 & -1.79 & -1.96 &  -2.29 & -2.19 \\
    \hline
        Median  &       & -2.46(4) & -2.44(1) &-2.33(6) & -2.22(1)\\
    \hline
    \end{tabular}
    \begin{tablenotes}
        \item \textit{Notes}. Medians were calculated after removing outliers. Lines with ``--'' were not well fit, with {\sf TSFitPy} marking them with flag\_error or flag\_warning greater than 0.
    \end{tablenotes}
    \end{threeparttable}
\end{table}

\subsection{Chemical abundances}

Chemical abundances were determined by spectral synthesis. The order for the determination of abundances followed the periodic table (ascending Z), except for the carbon abundance, which was the first to be determined. To account for possible blends because of the enhanced lines of heavy elements, the abundance determination loop was run twice. In the second time, all abundance values found in the first run were used as input values. We generated grids of spectra using {\sf TSFitPy} for each spectral line. The grids were generated with values from [X/Fe] = $-3.0$~dex to [X/Fe] = +3.0~dex in steps of 0.01 and $\chi^2$ values were calculated to compare the synthesized spectra and the observed spectrum. Then, a Nelder-Mead optimization algorithm, implemented using {\sf scipy} \citep{Virtanen2020SciPyPython}, was used to find the minimum. This approach was performed in a region of 1~nm around each spectral line. The final mean abundances are given in Tab.\ \ref{tab:chemicalabundances} and the line-by-line values are available in the data availability Section \ref{ap:abun}. For elements with multiple spectral lines, the abundance results are the mean with outliers excluded using a 3-$\sigma$ clipping criterion. Statistical uncertainties are the standard deviation of the line-by-line values. Some selected lines with their respective synthetic fits are shown in appendix \ref{ap:plots}.

The grid procedure was also applied to estimate the uncertainties of the abundances caused by the errors in the atmospheric parameters. This was done by changing the atmospheric parameters (\teff, \logg, \FeH~ and $\nu_{mic}$) up and down by their respective uncertainties and synthesizing spectra spanning 1~dex around the measured value of the abundance in steps of 0.02~dex. These uncertainties are also listed in Tab.\ \ref{tab:chemicalabundances}. 

Most of the atomic data we used are those compiled by the \Gaia-ESO Survey \citep{Heiter2021AtomicSurvey} complemented with data taken from the Vienna Atomic Line Database \citep[Vienna Atomic Line Database,][]{Ryabchikova2015ADatabase} below 420 nm. In the latter case, the lines are part of the line list compiled and tested as described in \citet{Giribaldi2023BerylliumSpectrograph}. When available, we adopted the hyperfine structure (HFS) of the spectral lines in the blue ($\leq$ 420 nm) from the {\sf linemake}\footnote{\url{https://github.com/vmplacco/linemake}} list generator \citep{Placco2021Linemake:Generator}. For lines in the redder part of the spectra ($\geq$ 420 nm), the \Gaia-ESO \citep{Heiter2021AtomicSurvey} HFS data was used. Exceptions to the above are the HFS for the lines of \ion{Eu}{II} and \ion{Ba}{II} that we re-calculated here. In summary, the elements Sc, V, Mn, Co, Nb, Ag, Ba, La, Pr, Nd, Sm, Eu, Tb, Ho, Yb, Lu, Ir, and Pb have HFS taken into account during the spectral synthesis.

For the Eu lines in the blue region (at 3724.9, 3819.7, 3907.1, 3930.5, 4129.7, and 4205.0 \AA) and the line at 6645.1 \AA\ in the red, we re-calculated the HFS taking into account isotopic splitting (isotopes with A= 151 and 153) adopting the data from \citet{Lawler2001ImprovedComposition}. For the Ba lines at 5853.67, 6141.71, and 6496.90 \AA, the calculations used the isotopic splitting and HFS data from \citet{VanHove1982J-dependentII} and \citet{Villemoes1993IsotopeSpectroscopy}. Hyperfine splitting is only important for isotopes with A = 135 and 137, but lines for isotopes with A = 134, 136, and 138 are also taken into account, as they are present in important amounts. The energy shifts and line strengths were calculated as in \citet{Smiljanic2023DetectingCUBES} with the help of the code made public by \citet{Mathar2011Corrigendum3259} and equations that can be found, for example, in \citet{Emery2006HyperfineStructure}. The data for the Eu lines can be found in Appendix \ref{ap:eu} and for the Ba lines in Appendix \ref{ap:ba}.

When possible, abundances were corrected for non-LTE effects adopting atomic models and lists already available in {\sf TSFitPy} \citep{Gerber2023Non-LTETurbospectrum}. The list of elements and references for the corrections is as follows: Al and Na \citep{Ezzeddine2018AnCalculations} and \citep{Nordlander2017Non-LTEStars}, Y \citep{Storm2023ObservationalClock}, Sr \citep{Bergemann2012NLTEData}, Ba \citep{Gallagher2019ObservationalAtmosphere}, Ni \citep{Bergemann2021SolarAbundance}, Ti \citep{Bergemann2011IonizationStars}, Co \citep{Bergemann2009NLTEConstants}, Fe \citep{Bergemann2012NLTEData}, Mn \citep{Bergemann2019ObservationalStars}, Mg \citep{Bergemann2017Non-localStars} and Ca \citep{Mashonkina2017InfluenceStars}.

\begin{table*}[]
    \centering
    \begin{threeparttable}
    \caption{Atmospheric parameters of the stars}
    \label{tab:atmosphericparameters}
    \begin{tabular}{ccccc}
    \hline
    \hline
         Star & \teff & \logg & \FeH$_{LTE}$ & $\nu_{mic}$\\
         Designation & (K) & (dex) & (dex) & (km/s) \\
    \hline
         BPS CS 29529-0089 & 5014(56) & 2.16(1) & -2.46 $\pm$ 0.04 (stat) $\pm$ 0.08 (par) & 1.5 \\
         TYC 9219-2422-1 &  5367(67) & 3.00(1) & -2.33 $\pm$ 0.06 (stat) $\pm$ 0.06 (par) & 1.5\\
         HD 122563 & 4615(69)$\dagger$ & 1.09(15)$\dagger$ & -2.593(92)$\ddagger$ & 1.2 $\ddagger$ \\
    \hline
    \end{tabular}
    \begin{tablenotes}
        \item \textit{Notes}. $\dagger$ Values adopted from \citet{Giribaldi2023TITANSStars}. $\ddagger$ Values from Giribaldi et al. (in prep.).
    \end{tablenotes}
    \end{threeparttable} 
\end{table*}

\begin{table*}[]
    \centering
    \begin{threeparttable}
    \caption{Positions and kinematics of the target stars}  
    \label{tab:positionsandkinematics}
    \begin{tabular}{ccccccc}
    \hline
    \hline
         Star & $\alpha$ & $\delta$ & $\rho$ & $\mu^*_\alpha$ & $\mu_\delta$ & $\varpi$\\
         Designation & ($^\circ$) & ($^\circ$) & (km/s) & (mas/a) & (mas/a) & mas\\
    \hline
         BPS CS 29529-0089 & 63.567768189(3) & -60.151309948(3) & 84.6(3) & 8.9446(158) & 4.0875(160) & 	0.2766(124) \\
         TYC 9219-2422-1 &  159.093941426(3) & -71.502865743(3) & 221.7(1) & 65.7952(140) & 	-29.7024(120) & 1.5552(113) \\
         HD 122563 & 210.631835043(7) & 9.685782710(6) & -26.13(4) & -189.5386(310) & -70.4150(250) & 3.0991(332) \\
    \hline
    \end{tabular}
    \begin{tablenotes}
        \item \textit{Notes}. Kinematic data from \citet{GaiaCollaboration2021GaiaProperties}. $\rho$ is the mean and the mean standard deviation for values in Tab. \ref{tab:logbook}. For HD 122563 radial velocity we adopted value from \citet{Deka-Szymankiewicz2018TheSample}.
    \end{tablenotes}
    \end{threeparttable}
\end{table*}

\subsection{Dynamics}

We integrated the stellar orbits using Gaia DR3 data \citep[coordinates, parallaxes and proper motions,][]{GaiaCollaboration2021GaiaProperties}. We used the Python code galpy \citep{Bovy2015Galpy:Dynamics} to integrate the orbits assuming the \citet{McMillan2017TheWay} Milky Way potential. The integration lasted for 13 Ga\footnote{The lower-case letter `a' is the symbol adopted by the International Astronomical Union (IAU) and the bureau of weights and measures of the International System of Units (BIPM/SI) for `year', and is thus the symbol used here.} backward in steps of 260 Ma. A Monte Carlo simulation was performed on the uncertainties of the positions, velocities, and parallaxes. We analyzed these stars in a similar manner as in \citet{DaSilva2023ExploringPopulations}.

\section{Results}
\label{sec:results}

As mentioned in Section \ref{sec:Methods}, we selected a sample of GALAH stars ignoring the quality flags of the Ba and Eu abundances. However, it turns out that the \BaFe~ in both \TYC~ and \BPSCS, selected for this pilot study, have flag 0 (indicating reliable values) in GALAH DR3. The abundances of Ba for the two stars in GALAH agree well with the values determined here. The \EuFe~for \BPSCS~is also very close to ours. These similarities are somewhat surprising, as the GALAH effective temperatures differ from our values by about 2~$\sigma$. In GALAH DR3, \teff~values were derived spectroscopically through the Fe excitation and ionization equilibria. GALAH also provide alternative \teff~values based on the IRFM from \citet{Casagrande2021TheSystem} and those values differ from ours by only 11~K. As for \EuFe~in \TYC, GALAH only provides an upper limit (indicated by a flag 1 in the GALAH tables) of +2.34(14), which is much higher than the value derived here.

\subsection{Light elements and Fe-peak}

We present the chemical abundances in Tab. \ref{tab:chemicalabundances}. The three stars analyzed in this study have low carbon abundances ([C/Fe]$<1.0$) and are therefore not part of any CEMP subclass of metal-poor star. Carbon corrections based on the evolutionary status of the stars \citep{Placco2014Carbon-EnhancedAbundances}\footnote{Using calculator available in \url{http://vplacco.pythonanywhere.com/}} for \BPSCS~ and \TYC~are on the same order of magnitude of the abundance uncertainties. However, for \HD~the carbon correction is 0.74, bringing its actual [C/Fe] ratio to about 0.06~dex. 

We tried to derive the abundances of N using the CN molecular bands between 387.5 and 390.0 nm, however those CN lines were found to be too faint. The forbidden oxygen line at 630.03 nm in both \BPSCS~ and \TYC~are unfortunately contaminated by a telluric line. 

The Li abundances were measured using the doublet lines around 670.7 nm. \BPSCS~ presents A(Li)=1.23 and \TYC~ A(Li)=1.29 while \HD~ A(Li)=$-$0.28. These values put both \BPSCS~and \TYC~at the Li plateau of metal-poor red giants discovered by \citet{Mucciarelli2022DiscoveryStars}, which is consistent with the Spite plateau \citep{Spite1982AbundanceConsequences.} of lithium measured in metal-poor dwarfs after the dilution by evolutionary effects is taken into account.

The $\alpha$ elements (Mg, Si, and Ca) are enhanced in all three stars, as expected, in particular after non-LTE corrections are taken into account (see Ca for \HD). 
Aluminum in the three stars is well below the solar [Al/Fe] ratio, with \TYC~ and \BPSCS~presenting [Al/Fe]$<-0.7$ dex. We also find that the three stars have low manganese abundances ([Mn/Fe] $<$ 0.0 dex). We refrain from discussing the stars in the [Mg/Mn] vs. [Al/Fe] plane, as an indicator of accreted origin, as this has been mostly used for stars with [Fe/H] $>~-$2.0 \citep{Das2020AgesStars}. 

Apart from Mn, the other iron peak elements between Sc and Zn in \BPSCS~mostly exhibit [Element/Fe] ratios close to solar. The only exception is Cr, which is found to be quite low at [Cr/Fe] = $-$0.51. Interestingly, \TYC~tends to have [Element/Fe] ratios that are 0.1 to 0.2 dex higher than \BPSCS~in all these elements. An even-odd effect can be observed in the abundances from C to Zn in both \BPSCS~ and \TYC.

For \HD, the abundance levels are all below solar. Abundances of iron-peak elements for \HD~have been determined, for example, by \citet{Barbuy2003Oxygen122563}. Our [Element/Fe] ratios are, on average, $-$0.13 dex lower than the values obtained by \citet{Barbuy2003Oxygen122563}. However, their metallicity value ([Fe/H] = $-$2.71) is lower than the one we adopted by 0.12 dex. This difference essentially fully explains the difference in [Element/Fe] ratios.

\subsection{Neutron capture elements}

Our results show that the \Gaia-Benchmark \HD~ presents no enrichment in neutron capture elements. This is in agreement with previous results in the literature \citep{Honda2006NeutronCapture122563}. The [Y/Fe] we determined in LTE ([Y/Fe] = $-$0.59) is actually within the interval presented by the NLTE analysis of \citet{Storm2023ObservationalClock} (see bottom panel of Fig. 8 of their paper).

\begin{figure}
    \centering
    \includegraphics[width=\linewidth]{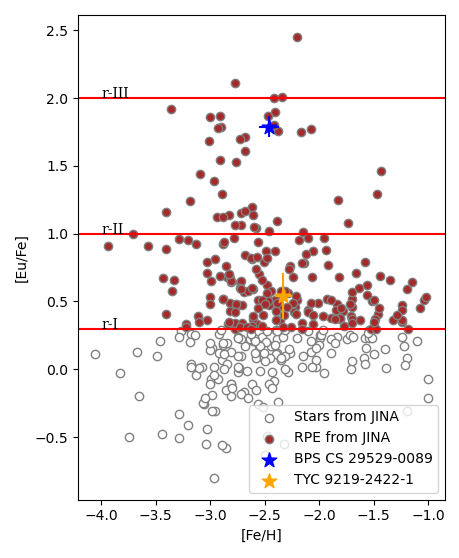}
    \caption{The \EuFe~ratio as a function of \FeH. The stars shown as circles are non-CEMP metal-poor stars from the JINA database \citep{Abohalima2018JINAbaseAStars}. Open circles are ordinary stars, while brown are RPEs. R-I, r-II and r-III levels are shown by the red lines. \BPSCS~and \TYC~are shown as colored star symbols.}
    \label{fig:EuFe_FeH}
\end{figure}

\TYC~presents a moderate but clear enrichment in neutron capture elements. Our new measurement of \EuFe=0.54(12) together with [Ba/Eu]=$-$0.70 (in LTE) indicate that \TYC~is an r-I star. \BPSCS~is found to have a stronger enrichment in neutron-capture elements. Star \BPSCS~ has \EuFe=1.79 and \BaEu=$-$0.66 (LTE). \BPSCS~ is thus a r-II star and the abundances point to a probable r-process only enrichment \citep[see][]{Bisterzo2014GalacticStructure}. We compare the \EuFe~ levels of \BPSCS~ and \TYC~ with 379 non-CEMP stars from the JINA database\footnote{Available in \url{https://jinabase.pythonanywhere.com/}} \citep{Abohalima2018JINAbaseAStars} in the same manner as \citet{Cain2020TheR-IIIStars}, see Fig.~\ref{fig:EuFe_FeH}. 

We detected Th in both \BPSCS~and \TYC, but the lines of uranium could not be detected. The high abundance of Th indicates an actinide-rich environment forming  \BPSCS. A fraction of metal-poor stars, mostly below [Fe/H]=$-$3.0, have been found to be enhanced in Th and U with respect to lighter r-process elements and have been called actinide-boost stars \citep[e.g.][]{Holmbeck2018TheKnown, Placco2023GHOST}. For \BPSCS~ we found $\log\epsilon~\mathrm{Th/Eu}$=$-$0.40 which, although above solar, implies that this star is not part of the actinide-boost class, even though we do not know its uranium abundance. 

\section{Discussion}\label{sec:discussion}

\subsection{Chemodynamic analysis}

\begin{figure}
    \centering
    \includegraphics[width=\linewidth]{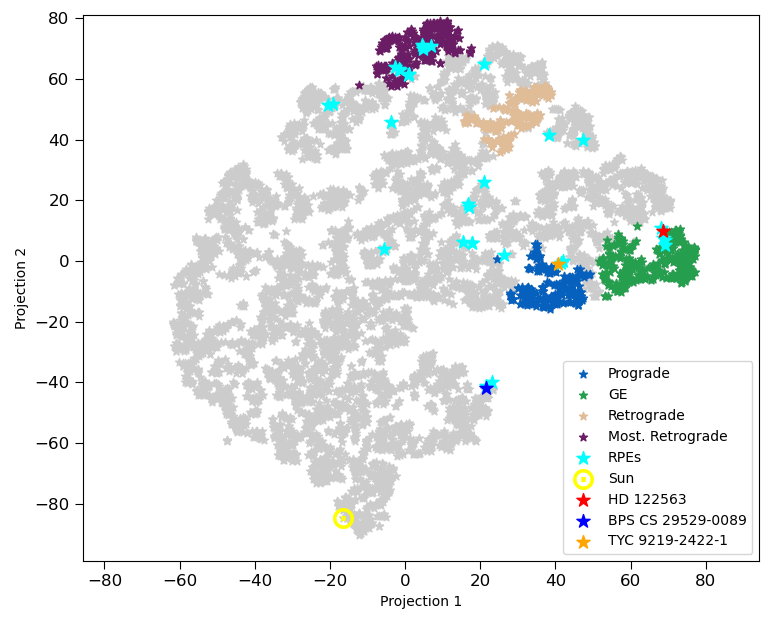}
    \caption{Projection map obtained using t-SNE. The gray stars are the sample from GALAH used in \citet{DaSilva2023ExploringPopulations}. The four groups defined in that work (Gaia Enceladus, prograde, retrograde, and most retrograde) are colored in green, blue, light brown, and dark purple, respectively. The cyan stars are 28 RPE stars with \EuFe$\geq$+1.7 dex found in the SAGA database \citep{Soubiran2023GaiaVersion} complemented by five stars from \citet{Roederer2024The+2.45}. The Sun is added as reference.}
    \label{fig:tsnemap}
\end{figure}

\begin{figure}
    \centering
    \includegraphics[width=\linewidth]{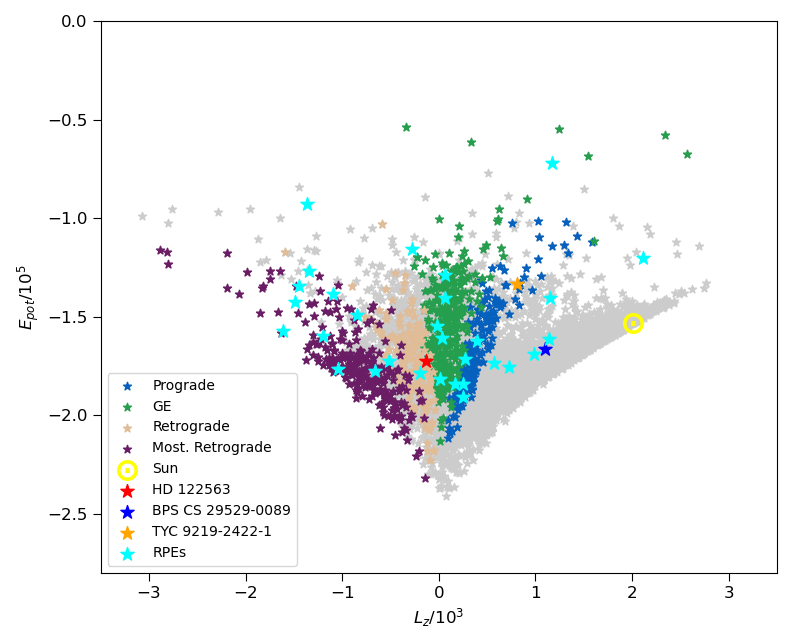}
    \caption{Lindblad Diagram (\Lz$\times$\En). Coloring is the same as Fig.~\ref{fig:tsnemap}. Error bars are smaller than the size of the symbols.}
    \label{fig:lindblad}
\end{figure}

We used the dimensionality reduction algorithm t-SNE (t-distributed stochastic neighbour embedding) for the chemodynamic classification of the stars, in a similar manner as done in \citet{DaSilva2023ExploringPopulations}. For this analysis, we adopted four quantities, \FeH, \MgFe, \Jx~ and \Jy\footnote{\Jx~ and \Jy~ are the axis of the Action Map, we use a system that \Jphi=\Lz.}, as dimensions, and use t-SNE to obtain a projection of the stellar distribution in two dimensions (2D). The 2D projection is shown in Fig.~\ref{fig:tsnemap}. To perform the classification, we also included the stars of the GALAH sample used in \citet{DaSilva2023ExploringPopulations}. The same four groups defined in that work are highlighted in Fig.~\ref{fig:tsnemap}, i.e., a group with average zero angular momentum associated to Gaia Enceladus, a second group that is slightly more prograde, a third group that is slightly more retrograde, and a fourth group that includes the most retrograde stars of the sample (see Fig. \ref{fig:lindblad}). We recall that \citet{DaSilva2023ExploringPopulations} found that the Gaia Enceladus group was chemically similar to both the retrograde and prograde groups. Only the most retrograde group showed distinct chemical composition and was suggested to be dominated by stars from the Sequoia and/or Thamnos mergers \citep[see][]{Myeong2019EvidenceHalo, Koppelman2019MultipleHistory}. For comparison purposes, we also added to the sample several RPEs from the SAGA database and from \citet{Roederer2024The+2.45} that have \EuFe$\geq+1.70$ dex.

\begin{figure}
    \centering
    \includegraphics[width=\linewidth]{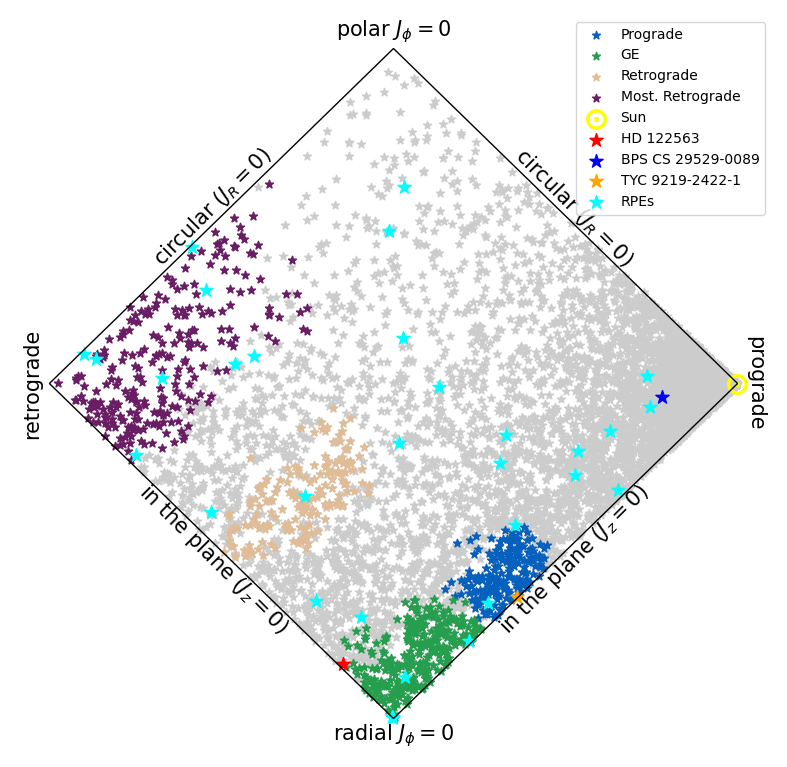}
    \caption{Action map. Coloring is the same as Fig.~\ref{fig:tsnemap}. Note \BPSCS~ in the most prograde, circular and in the plane orbits.  Error bars are smaller than the size of the symbols.}
    \label{fig:actionmap}
\end{figure}

In the tSNE projection, \BPSCS~falls in the region dominated by disk stars (bottom part, with values below $\sim$ $-$20 in the projetion 2 axis). Both \HD~and \TYC~fall in the other region, dominated by halo objects. The action map shown in Fig.\ \ref{fig:actionmap} supports this, as \BPSCS~is seen to have an orbit that is prograde, circular, and in the plane. Compared to the action map of \citet{Shank2023TheStars}, \BPSCS~fits their definition of the metal-weak thick disk (MWTD) group. These stars are thought to be primordial proto-disk stars that have migrated due to interactions within the Milky Way or with merger events\citep{Shank2023TheStars}. Interestingly, \cite{Shank2023TheStars} indicates that the MWTD does not seem to have suffered r-process enrichment events that would create a significant fraction of RPE stars \citep{Shank2022DynamicallySurvey, Shank2022Dynamically6}. This already highlights the uniqueness of \BPSCS. 

Three other RPE stars fall in the same region as \BPSCS~ in the t-SNE map: HE 1105+0027, TYC 7325-920-1 and CD-28 1082. However, only two of them (HE 1105+0027 and CD-28 1082) remain close to \BPSCS~in the action map (Fig.~\ref{fig:actionmap}) and in the Lindblad diagram (Fig.~\ref{fig:lindblad}). TYC 7325-920-1, with \FeH = $-$2.80, [C/Fe] = 0.56, \BaFe = 1.35, and \EuFe = 2.23 \citep{Cain2020TheR-IIIStars}, has a high \zmax~ and kinematic and does not match the thick disk (see Tab.~\ref{tab:kinematics}). 

However, CD-28 1082 and HE 1105+0027 have thick disk characteristics. CD-28 1082 (\FeH = $-$2.45) is highly enhanced in C, Ba, and Eu; [C/Fe] = 2.19, \BaFe = 2.09, and \EuFe = 2.07 \citep{Purandardas2019ChemicalAbundances,Goswami2021SpectroscopicNucleosynthesis}. HE 1105+0027 (\FeH = $-$2.42) is similarly enhanced in C, Ba, and Eu; [C/Fe] = 2.00, \BaFe = 2.45, and \EuFe = 1.81 \citep{Goswami2021SpectroscopicNucleosynthesis}. These stars have dynamic properties similar to \BPSCS~. This can be easily seen in the Lindblad diagram (Fig.~\ref{fig:lindblad}) and the action map (Fig.~\ref{fig:actionmap}), which show only these two stars nearby \BPSCS. The characteristic that distinguishes \BPSCS~is that it has a normal carbon abundance. This star seems to be the only known very metal-poor r-II star belonging to the old disk.

Our second target, \TYC~falls on the prograde group that \citet{DaSilva2023ExploringPopulations} associated to \Gaia-Sausage-Enceladus \citep[see e.g.][]{Helmi2018TheDisk, Belokurov2018Co-formationHalo}. \citet{Aguado2021ElevatedSequoia} and \citet{Matsuno2021R-processDR3} report a \BaFe$\leq$ 0.0 and \EuFe$\sim$ +0.5 which is in good agreement with \TYC. \citet{DaSilva2023ExploringPopulations} showed that the \Gaia-Sausage-Enceladus stars have [Eu/Mg] $\sim$ 0.40 in the metallicity interval between about $-$0.8 and $-$1.7. More recently, \citet{Monty2024TheClusters}, \citet{Ou2024TheGalaxy} and \citet{Ernandes2024Gaia-Sausage-EnceladusAbundances} found that the [Eu/$\alpha$] ratio decreases for metallicities below that interval, reaching [Eu/$\alpha$] $\sim$ 0.0 (with some scatter) at the metallicity of \TYC. As discussed in \citet{Ou2024TheGalaxy}, for example, delayed r-process sources, such as NSMs, are needed to explain the increase in the [Eu/Mg] ratio with increasing metallicity. We compare \TYC~ with stars in \citet{Ou2024TheGalaxy} in Fig.~\ref{fig:TYC_MgH_EuMg}. Its position on the [Mg/H] by [Eu/Mg] plane matches the position of the other stars in the magnesium-poor tail of their sample. 

In our analysis, \HD~ falls on \Gaia-Sausage-Enceladus group proper (the one with average zero angular momentum in Fig. \ref{fig:lindblad}). However, it terms of its chemistry, besides \FeH~ and \MgFe~values that agree with \Gaia-Sausage-Enceladus, \HD~is extremely poor in neutron capture elements. Therefore, we suggest that \HD~is instead part of the in situ population that contaminates this region of the parameter space \citep[as discussed in][in situ populations seem to be present in all groups that were investigated in that work]{DaSilva2023ExploringPopulations}.

\subsection{\BPSCS~as part of the proto-disk population}

As mentioned in the Introduction, RPEs have been shown to generally be part of the halo stellar population. As such, they are either accreted stars from dwarf galaxies that merged with the Milky Way, from systems maybe similar to ultra-faint dwarf galaxies (UFD) or dwarf spheroidals (dSph) \citep[see e.g.][]{Ji2016COMPLETEII, Reichert2020Neutron-captureGalaxies,Reichert2021ExtremeFornax}, or they are part of the \textit{in-situ} halo of our Galaxy. In support of the idea of accreted origin, RPEs have been found in streams such as Indus \citep{Hansen2021Stream} or in dwarf galaxies such as Fornax \citep{Ji2016COMPLETEII} and Reticulum II \citep{Reichert2020Neutron-captureGalaxies}. Some studies have shown that a fraction ($\sim$ 10\%) of RPE stars with origin in the Milky Way display disk dynamics \citep[see e.g][]{Gudin2021TheHistory, Shank2022DynamicallySurvey}. More recently, the first thin-disk RPE star (with [Fe/H] $>$ $-$1.0) has been discovered \citep{Xie2024Discovery+0.78}.

Cosmological zoom-in simulations of Milky Way-like galaxies could shed light on where RPE form. In this context, \citet{Hirai2022OriginGalaxy, Hirai2024TheGalaxy} simulated the formation of small stellar systems that were later accreted into a Milky Way-like galaxy. Both studies concluded that RPE stars are more easily formed in small dwarf systems. This is due to the fact that a single Neutron Star Merger (NSM) can significantly increase the abundance of r-process elements in these systems because of their relatively low gas mass. Observationally, \citet{Limberg2024ExtendingWukong/LMS-1} note that on these small systems a delayed source is enough to explain the r-process enrichment. 

\citet{Hirai2022OriginGalaxy, Hirai2024TheGalaxy} also found that in-situ r-process enriched stars can be formed. This occurs in clumps contaminated with NSM ejecta in which mixing has not had enough time to dilute the enrichment before new stars are formed. \citet{Hirai2022OriginGalaxy} points that around 90\% of r-II stars with \FeH$\leq -2$ in a Milky-Way like galaxy have accreted origins. In the newer study, they point out that less than 20\% of stars with $+1.5\leq$\EuFe$\leq +1.9$ would have \textit{in-situ} origin. Moreover, \citet{Hirai2024TheGalaxy} found that only 0.2\% of r-II RPE stars have origin in the \textit{in-situ} component of their simulation. Together, the results discussed above show that there is a plausible scenario to form an r-II star on the Milky Way disk and confirm that \BPSCS~ is an example of a very rare object. 

\subsection{Stellar abundances and the origin of their r-process elements}

Complete inventories of chemical abundances in metal-poor stars are somewhat rare in the literature. Some examples of RPE with more than 35 elemental abundances derived are: HD 222925 \citep{Roederer2022The}, 2MASS J22132050-513785 \citep{Roederer2024The+2.45}, 2MASS J00512646-1053170 \citep{Shah2024TheSpectroscopy}, HD~20 \citep{Hanke2020A20} and BPS CS  31082-001 \citep{Hill2002TheCosmochronology,Ernandes2022BeSpectroscopy, Ernandes2023Reanalysis31082-001}. For our target \BPSCS, we provide abundances for 44 elements, of which 27 are neutron-capture elements. This is a slightly higher number of elements than those found for HD 222925 \citep{Roederer2022The} using only ground UV and the visible part of the spectrum. Thus, \BPSCS~ is an excellent candidate for space UV observations, $\leq 3000$~\AA, to increase the number of neutron-capture elements with abundances, possibly including some rare elements like gold.

\begin{figure}[!ht]
    \centering
    \includegraphics[width=\linewidth]{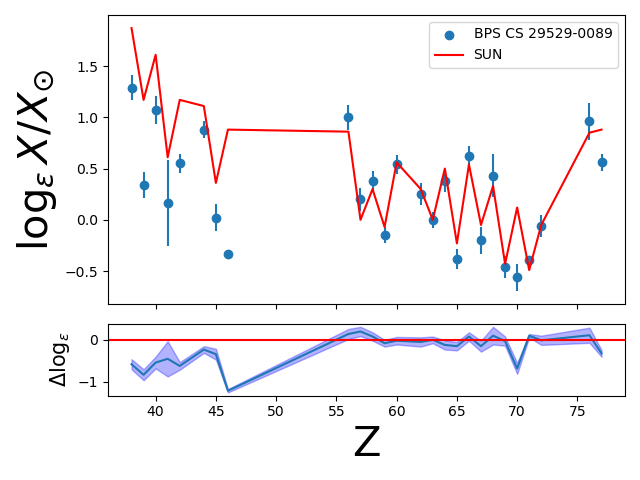}\\
    \includegraphics[width=\linewidth]{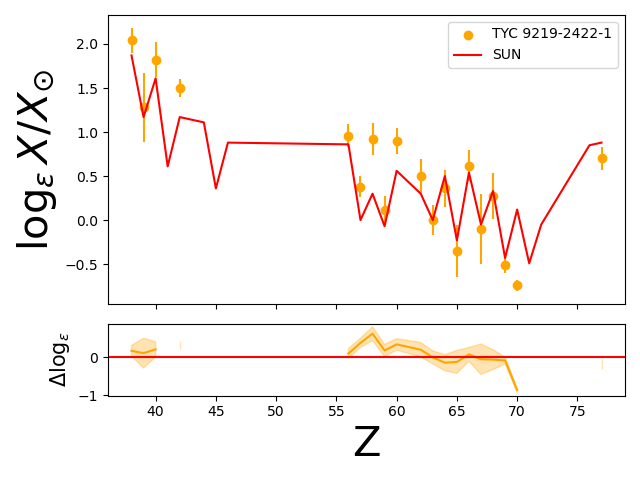}
    \caption{\textit{Top} - Comparison of the solar r-process abundance pattern with the abundances of \BPSCS. \textit{Bottom} - Same comparison for \TYC.}
    \label{fig:solarrprocess}
\end{figure}

\begin{figure}
    \centering
    \includegraphics[width=\linewidth]{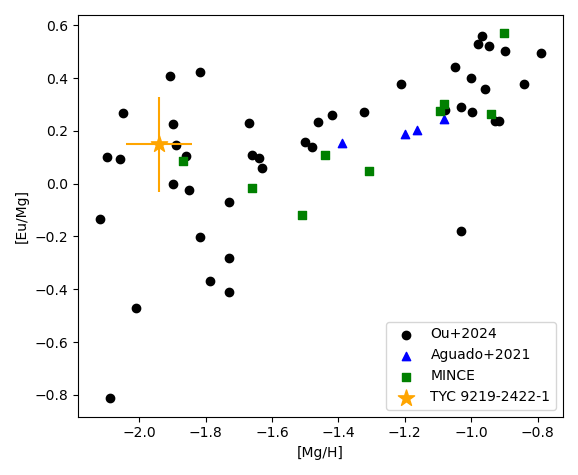}
    \caption{The [Mg/H] by [Eu/Mg] plane. Black circles are stars from \citet{Ou2024TheGalaxy}, green squares are stars from MINCE \citep[\textit{as cited in} \citeauthor{Ou2024TheGalaxy} \citeyear{Ou2024TheGalaxy}]{Cescutti2022MINCESample,Francois2024MINCEElements}, and blue triangles are stars from \citet{Aguado2021ElevatedSequoia}. \TYC~is the orange star.}
    \label{fig:TYC_MgH_EuMg}
\end{figure}

In Fig.~\ref{fig:solarrprocess}, we compare our results with the solar r-process abundance pattern. To obtain the solar r-process component, we subtract the expected contribution of the s-process using the fractions from \citet{Bisterzo2014GalacticStructure}. In the comparison, we normalise the abundance pattern to the Eu abundance. For \BPSCS, the most significant deviations occur for elements with Z $<$ 50. Such deviations have been seen before in other stars and have been attributed to the contribution of a weak r-process (or similar mechanism) to the solar material. This process is believed to synthesise lighter elements such as Sr (Z = 38), Y (Z = 39), and Zr (Z = 40), but it is not able to produce heavier than the first-peak r-process elements \citep{SiqueiraMello2014High-resolutionStars, Xing2024DetectionOrigin}. In contrast, \TYC~ appears to follow the solar r-process pattern for these lighter elements more closely, but we see some discrepancy in the abundances of Ce (Z = 58) and Nd (Z = 60). Finally, for both stars, ytterbium (Yb, Z = 70) stands out as an element that does not align with the expected pattern.

To gain some insight into the possible origins of abundances in \BPSCS~ and \TYC, we used {\sf starfit}\footnote{\url{https://starfit.org/}}\citep{Heger2010NucleosynthesisStars}. The code {\sf starfit} compares the measured abundance pattern in a star to a series of model abundances from different possible progenitors. Here we used the option with a genetic algorithm to choose the best fitting model(s) among the calculations of \citet{Heger2010NucleosynthesisStars}, \citet{Mendoza-Temis2015NuclearMergers}, and \citet{Wu2016ProductionMergers}. 

Our choice of models for {\sf starfit} means that we are searching only for solutions that combine one Population III (Pop III) core collapse supernova (SN) with a second source responsible for the r-process. This does not need to be the correct combination of sources needed to explain the abundances in our two stars. In fact, given their relatively high metallicity, [Fe/H] = $-$2.3 to $-$2.5, we should anticipate that there is a low probability that the stars are direct descendants of Pop III SNe. Nevertheless, there are theoretical simulations that indicate that a single Pop III SN could enrich the surrounding medium to a level close to [Fe/H] = $-$2.0 \citep[e.g.][]{Smith2015TheMini-haloes}. Moreover, the work of \citet{Hartwig2018DescendantsStars} shows that other chemical abundances, in addition to Fe, are important in the search for such mono-enriched stars. These authors investigated regions of the chemical abundance parameter space with high probability of being populated by second-generation stars enriched by a single Pop III supernova.

Following \citet{Hartwig2018DescendantsStars}, we looked at the abundances of C, Mg, Ni, Sc, Mn, and Al (in addition to Fe) compared to their chemical divergence maps (see their Fig. 18). For \BPSCS, we have [Mg/C] = 0.30(13) dex, [C/Ni] = 0.04(17) dex, and [Sc/Mn] = 0.39(16) dex, while for \TYC, the corresponding values are [Mg/C] = 0.27(14) dex, [C/Ni] = 0.07(23) dex, and [Sc/Mn] = 0.61(19) dex. These measurements place both \BPSCS~ and \TYC~ within the regions dominated by mono-enriched stars in Fig. 18 of \citet{Hartwig2018DescendantsStars}. However, we remark that, as discussed in \citet{Hartwig2018DescendantsStars}, this comparison does not give a probability that the stars are mono-enriched. It simply shows that in the chemical abundance space they are located in a region dominated by mono-enriched stars. There may be other means for such abundances to be obtained apart from a single Pop III star. In any case, this comparison indicates that it is interesting to use {\sf starfit} to search for Pop III progenitors that could explain our two stars.

\begin{figure*}
    \centering
    \includegraphics[width=\linewidth]{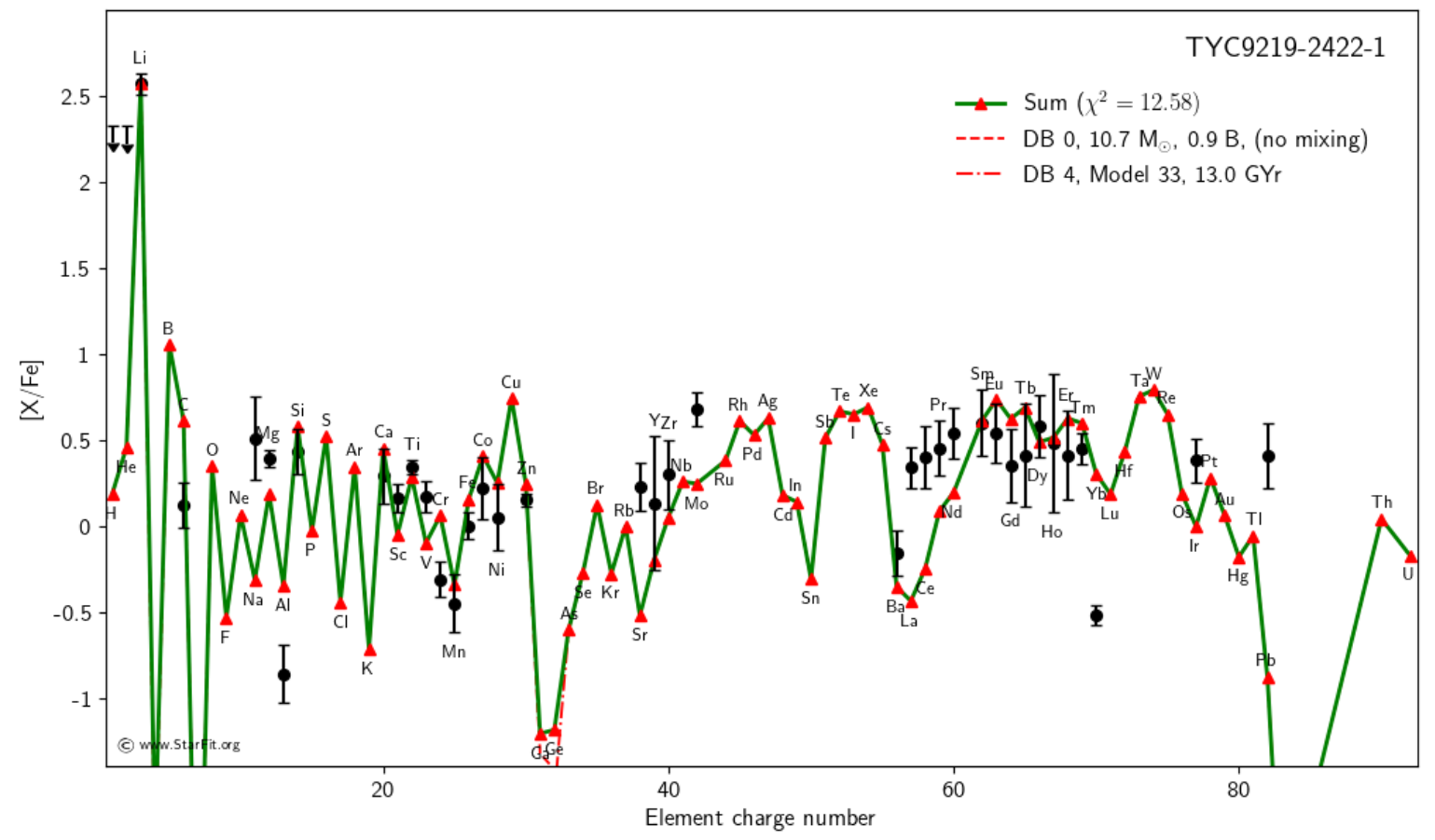}
    \caption{Model fit using starfit for the abundances of \TYC. The dashed line is a model from the DB0 database of Pop III progenitor stars from \citet{Heger2010NucleosynthesisStars} and later updates. The dashed dotted line is a model from the DB3 database of NSM r-process models from \citet{Mendoza-Temis2015NuclearMergers} and \citet{Wu2016ProductionMergers}. Model 33, that best fits the neutron-capture abundances here, is a NSM model which forms a black hole of about 3\Msun with a torus ejecta of 0.10\Msun\citep[for more details see model m0.10 in][]{Wu2016ProductionMergers}. Black points are the observed abundances. The green line with red triangles is the sum of the two best fitting models.}
    \label{fig:TYC_starfit}
\end{figure*}

A fit for the abundance pattern of \TYC~ is provided in Fig.~\ref{fig:TYC_starfit}. From this fit, the most probable phenomenon leading to the chemical abundance pattern of this star is a combination between one Pop III supernovae, with the progenitor mass of 10.7 \Msun~(fitting the lighter elements with the even-odd pattern until Zn), with a NSM r-process phenomenon occurring 13 Ga ago (fitting heavier elements from Sr upwards). If correct, this fit would agree with the findings of \citet{Ernandes2024Gaia-Sausage-EnceladusAbundances}, which point to an important enrichment of r-process elements coming from NSM in stars of Gaia-Sausage-Enceladus.

\begin{figure*}
    \centering
    \includegraphics[width=\linewidth]{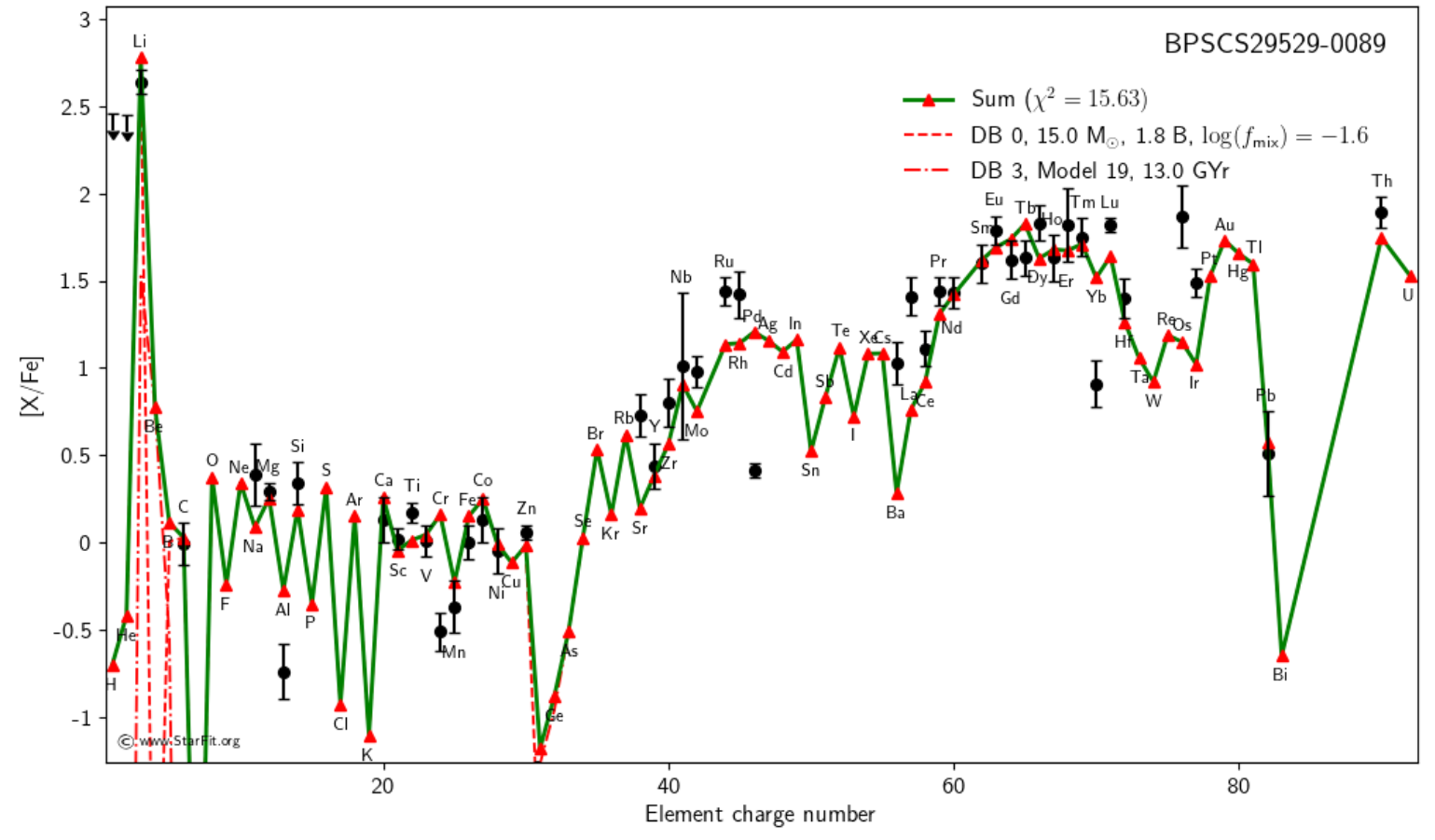}
    \caption{Model fit using starfit for the abundances for \BPSCS. The dashed line is a model from the DB0 database of Pop III progenitor stars from \citet{Heger2010NucleosynthesisStars} and later updates. The dashed dotted line is a model from the DB3 database of NSM r-process models from \citet{Mendoza-Temis2015NuclearMergers} and \citet{Wu2016ProductionMergers}. Model 19, that best fits the neutron-capture abundances here, is a NSM model which forms a black hole of about 3\Msun with a torus eject of 0.03\Msun\citep[for more details see the similar model S-def in][]{Wu2016ProductionMergers}}. Black points are the observed abundances. The green line with red triangles is the sum of the two best fitting models.
    \label{fig:BPS_starfit}
\end{figure*}

The abundance pattern for \BPSCS~ is shown in Fig.~\ref{fig:BPS_starfit} in conjunction with the best fitting models. The abundance pattern of \BPSCS~ indicates a combined contribution from a Pop III supernovae with 15 \Msun~ and a NSM r-process enrichment 13 Ga ago. The NSM model fits the abundances of this star better than what was seen with \TYC. A possible NSM origin for the r-process enrichment of \BPSCS~is consistent with the scenario proposed in \citet{Hirai2022OriginGalaxy, Hirai2024TheGalaxy} for the formation of such stars in-situ in the proto-disk of the Milky Way, as discussed above.

We note that both stars present an excess in the abundance of the elements around Z$\sim 40$ and $\sim 58$ with respect to the models. This could be explained with fission products, as suggested in \citet{Roederer2023ElementUranium}. Fission products are daughter nuclides of spontaneous fission that neutron-rich father nuclides undergo, especially for transuranic elements (with Z $\geq 92$). This stochastic process produces pairs of daughter nuclides with a two-peak distribution with maxima around A$\sim 95$ or $A\sim 140$. \citet{Roederer2023ElementUranium} argues that the strong correlation between the abundances of elements such as ruthenium, rhodium, palladium, and silver points to an excess abundance caused by fission fragments of transuranic elements with a large number of neutrons. In fact, the values of $\log\epsilon~\mathrm{(Ru/Zr)}$ = $-$0.19, $\log\epsilon~\mathrm{(Rh/Zr)}$ = $-$1.05 and $\log\epsilon~\mathrm{(Pd/Zr)}$ = $-$0.48 for \BPSCS, are similar to the expected values to be found in case of fission, around $-$0.18, $-$0.82 and $-$0.54, respectively, for the level of \EuFe seen in the star \citep[see Fig. 8 in][]{Roederer2024The+2.45}. 

As a last point, we note that essentially all heavier neutron capture abundances appear to follow the expected pattern, except for the element Yb. We rechecked its abundance, determined using the \ion{Yb}{ii} line in 3694\AA~for the three stars (see Fig.~\ref{fig:BPS_Yb} in Appendix \ref{ap:plots}). We confirmed the values, but we do not have a clear explanation for the apparent deficiency.


\subsection{Nucleocosmochronology}

Most of nuclides are unstable. They decay in a stochastic manner with the rate of decay as a function of time. The phenomenon of radioactive decay has been used to determine accurate and precise ages in a series of fields from archaeology to environmental radiometry. Although the use of radioactivity for the determination of age of stars has been suggested as a result of its convenience, there are limitations. Specifically, it is not feasible to directly measure radionuclide abundances; only total elemental abundances are accessible, and the initial production ratios remain highly uncertain \citep{Cayrel2001MeasurementDecay}.

Given the detection of both Th and Hf in \BPSCS, we decided to try to date this star using nucleocosmochronology. We follow here \citet{Kratz2007ExplorationsStars}, who suggested that the Th/Hf ratio is possibly one of the most accurate cosmochronometers, given that the production of Hf tends to closely follow the production of heavier elements of the third r-process peak. 

We used the equation in \citet{Cayrel2001MeasurementDecay} and initial ratios from both \citet{Schatz2002Thorium31082-001} and \citet{Kratz2007ExplorationsStars} to calculate the nucleocosmochronological age of \BPSCS. As usual in this type of calculation, a large scatter in values is found \citep[see also][]{Ludwig2010AccuracyStars}. The Th/Hf ratio itself, using the initial ratios from \citet{Kratz2007ExplorationsStars}, seems to give an age in agreement with what should be expected for such metal-poor stars, an estimated age of 12.91 $\pm$ 6.64(abund.) $\pm$ 1.03 (PR) Ga. 

\section{Conclusions}
\label{sec:conclusions}     

In this work, we present the chemical abundance analysis of \BPSCS, \TYC~,and \HD~ using high-resolution spectra obtained with the UVES spectrograph. Atmospheric parameters were determined iteratively using IRFM calibrations, bolometric corrections, \Gaia~ parallaxes and \ion{Fe}{ii} lines. We also determined abundances for 47 chemical species of 44 elements through spectral grid generation. Corrections for departures from LTE were used for 15 chemical species of 13 elements. We investigated the chemodynamics of the stars following the procedure outlined in \citet{DaSilva2023ExploringPopulations}, to investigate the possible origins of the objects in the Milky Way evolution context. We find that their abundances of C, Mg, Ni, Sc, Mn, and Al are consistent with expectations for mono-enriched stars, following \citet[][see their Fig. 18]{Hartwig2018DescendantsStars}.

\BPSCS~has enhanced abundances of a series of elements produced by neutron-capture processes. This made possible the determination of an almost complete set of heavy elements with lines available between 330 nm to 670 nm.
We found that \BPSCS~is the first known r-II star (\EuFe$=+1.79$ dex) with thick disk dynamics that is not carbon enhanced. Two other r-II stars in the metal-weak thick disk are known, but they are both CEMP. We suggest that the most probably \BPSCS~was formed \textit{in situ}, in the proto-disk of the Milky Way. Alongside other thick disk stars, its kinematics was probably heated during the merger events suffered by the Milky Way, such as the one with \Gaia-Sausage-Enceladus. The abundance pattern of \BPSCS~could be explained by the combination of two phenomena: i) for elements with Z $\leq$ 30 (until Zn), a supernova of a Pop III progenitor with 15 \Msun, and ii) for elements with Z $\geq$ 38 (from Sr), a NSM r-process enrichment happening about 13 Ga ago. We speculate that fission products could explain some of the excesses of neutron-capture elements seen around Z $\sim$ 40 and 58, in the same way as discussed in \citet{Roederer2024The+2.45}.

\TYC~is an archetypal star from \Gaia-Sausage-Enceladus. It has a moderate europium enrichment of \EuFe$=+0.54(17)$ dex with \BaFe$=-0.16(12)$, which leads to the r-I classification. The [Eu/Mg] ratio observed in \TYC~is also in agreement with the range observed for the \Gaia-Sausage-Enceladus magnesium-poor tail \citep{Matsuno2021R-processDR3,Ou2024TheGalaxy}. The orbit of \TYC, although slightly prograde, is highly eccentric $e\sim0.8$ which agrees with the \Gaia-Sausage-Enceladus well-known high eccentricity \citep[see e.g.,][]{Helmi2008TheGalaxy,Koppelman2020TheGalaxy,Limberg2021DynamicallySurveys,DaSilva2023ExploringPopulations}. \TYC abundance pattern could be explained by the combination of two phenomena: i) for elements with Z $\leq$ 30 (until Zn), a supernova of a Pop III progenitor with 10.7 \Msun, and ii) for elements with Z $\geq$ 38 (from Sr), a NSM r-process enrichment hapening about 13 Ga ago.

\HD~was in principle also found to associated to \Gaia-Sausage-Enceladus, based on the values of actions, [Fe/H], and [Mg/Fe]. However, this star has very low abundances of neutron capture elements. This pattern does not match the established \Gaia-Sausage-Enceladus r-process enrichment \citep{Aguado2021ElevatedSequoia,Matsuno2021R-processDR3}. Therefore, most probably \HD~is an old metal-poor in-situ halo star.

\section{Data Availability}
\label{ap:abun}
A list of lines, log gf, and abundances will are provided in \href{https://doi.org/10.5281/zenodo.14982574}{Zenodo}.

\begin{acknowledgements}
   ARS and RS acknowledge the support of the Polish National Science Centre (NCN) through the project 2018/31/B/ST9/01469. ARS thanks Riano Giribaldi for the discussion on HD122563 and the data provided. ARS thanks Hélio Perottoni, Maria L. L. Dantas and Guilherme Limberg for the discussions. This work is based on observations collected at the European Southern Observatory under ESO programmes:  0108.D-0626(A) and 65.L-0507. This research has made use of the SIMBAD database, operated at CDS, Strasbourg, France and of NASA’s Astrophysics Data System. This research has made use of the NASA/IPAC Infrared Science Archive, which is funded by the National Aeronautics and Space Administration and operated by the California Institute of Technology. This work has made use of data from the European Space Agency (ESA) mission
{\it Gaia} (\url{https://www.cosmos.esa.int/gaia}), processed by the {\it Gaia} Data Processing and Analysis Consortium (DPAC,
\url{https://www.cosmos.esa.int/web/gaia/dpac/consortium}). Funding for the DPAC has been provided by national institutions, in particular the institutions participating in the {\it Gaia} Multilateral Agreement. This work also made use of GALAH survey data, their website is \url{http://galah-survey.org/}. 

\end{acknowledgements}
    
\bibliographystyle{aa} 
\bibliography{references} 

\clearpage

\begin{appendix}
\section{Chemical abundances and Dynamics}
Tables with the determined chemical abundances and dynamic parameters are available in tables \ref{tab:chemicalabundances} and \ref{tab:kinematics}.
\begin{table*}[]
    \tiny
    \centering
    \begin{threeparttable}
    \caption{Determined Abundances and uncertainties}  
    \label{tab:chemicalabundances}
    \begin{tabular}{l|ccccc|ccccc|ccccc}
    \hline
    \hline
         Element & \multicolumn{5}{c|}{BPS CS 29529-0089} & \multicolumn{5}{c|}{TYC 9219-2422-1} & \multicolumn{5}{c}{HD 122563}\\
         & Abund. & $\sigma_{stat}$ & $\sigma_{par}$ & NLTE & \#lines & Abund. & $\sigma_{stat}$ & $\sigma_{par}$  & NLTE & \#lines & Abund. & $\sigma_{stat}$ & $\sigma_{par}$ & NLTE & \#lines\\
    \hline
         $\mathrm{[C/Fe]}$  &-0.01 & --   & 0.12 & --   & band & 0.12 & --   & 0.13 & --   & band &-0.68$\star$ & --   & 0.12 & --   & band\\
         $\mathrm{[Li/Fe]}$ & 2.64 & --   & 0.07 & --   & 2    & 2.57 & --   & 0.06 & --   & 2    & 1.26 & --   & 0.14 & --   & 2   \\
         $\mathrm{[Na/Fe]}$ & 0.39 & 0.15 & 0.10 & 0.40 & 2    & 0.51 & 0.22 & 0.10 & 0.25 & 2    & 0.16 & 0.11 & 0.28 &-0.23 & 2   \\
         $\mathrm{[Mg/Fe]}$ & 0.29 & 0.01 & 0.05 & 0.19 & 2    & 0.39 & 0.01 & 0.05 & 0.26 & 2    & 0.23 & 0.02 & 0.12 & 0.32 & 2   \\
         $\mathrm{[Al/Fe]}$ &-0.74 & 0.05 & 0.15 &-0.73$\dagger$ & 2    &-0.86 & 0.01 & 0.17 &-0.83$\dagger$& 2    &-0.44 & 0.04 & 0.25 &-0.64$\dagger$& 2   \\
         $\mathrm{[Si/Fe]}$ & 0.34 & 0.11 & 0.04 & 0.26 & 9    & 0.43 & 0.13 & 0.03 & 0.26 & 13   & 0.19 & 0.12 & 0.10 & 0.16 & 14  \\
         $\mathrm{[Ca/Fe]}$ & 0.13 & 0.12 & 0.05 & 0.13 & 38   & 0.29 & 0.15 & 0.06 & 0.15 & 37   &-0.02 & 0.12 & 0.13 & 0.32 & 38  \\
         $\mathrm{[Sc/Fe]}$ & 0.02 & 0.04 & 0.05 & --   & 8    & 0.16 & 0.07 & 0.04 & --   & 8    &-0.10 & 0.05 & 0.11 & --   & 8   \\
       $\mathrm{[Ti~I/Fe]}$ & 0.05 & 0.01 & 0.08 & 0.10 & 4    & 0.27 & 0.02 & 0.08 & 0.26 & 4    &-0.18 & 0.05 & 0.17 &-0.15 & 4   \\
      $\mathrm{[Ti~II/Fe]}$ & 0.17 & 0.05 & 0.04 & 0.30 & 2    & 0.34 & 0.02 & 0.04 & 0.37 & 2    &-0.01 & 0.04 & 0.11 & 0.48 & 2   \\
       $\mathrm{[V~I/Fe]}$  &-0.15 & 0.08 & 0.12 & --   & 2    & 0.03 & 0.06 & 0.11 & --   & 2    &-0.35 & 0.08 & 0.16 & --   & 2   \\
      $\mathrm{[V~II/Fe]}$  & 0.01 & 0.08 & 0.05 & --   & 10   & 0.17 & 0.08 & 0.04 & --   & 11   & 0.04 & 0.12 & 0.12 & --   & 10  \\
         $\mathrm{[Cr/Fe]}$ &-0.51 & 0.06 & 0.09 & --   & 10   &-0.31 & 0.04 & 0.09 & --   & 10   &-0.73 & 0.03 & 0.18 & --   & 10  \\
       $\mathrm{[Mn~I/Fe]}$ &-0.57 & 0.21 & 0.10 &-0.43 & 10   &-0.53 & 0.22 & 0.11 &-0.54 & 10   &-0.72 & 0.09 & 0.20 &-0.45 & 8   \\
      $\mathrm{[Mn~II/Fe]}$ &-0.37 & 0.12 & 0.09 &-0.42 & 8    &-0.45 & 0.15 & 0.08 &-0.46 & 8    &-0.10 & 0.07 & 0.19 &-0.37 & 7   \\
         $\mathrm{[Co/Fe]}$ & 0.13 & 0.09 & 0.09 & 0.20 & 19   & 0.22 & 0.13 & 0.12 & 0.18 & 19   & 0.07 & 0.07 & 0.25 & 0.11 & 11  \\
         $\mathrm{[Ni/Fe]}$ &-0.05 & 0.11 & 0.07 & 0.27 & 12   & 0.05 & 0.18 & 0.06 & 0.20 & 17   &-0.24 & 0.06 & 0.13 & 0.25 & 14  \\
         $\mathrm{[Cu/Fe]}$ & 0.00 & --   & --   & --   & 3    & 0.29 & --   & --   & --   & 1    & 0.50 & --   & --   & --   & 3   \\
         $\mathrm{[Zn/Fe]}$ & 0.06 & --   & 0.04 & --   & 1    & 0.15 & --   & 0.04 & --   & 1    &-0.06 & --   & 0.10 & --   & 1   \\
         $\mathrm{[Sr/Fe]}$ & 0.73 & 0.09 & 0.08 & 0.78 & 3    & 0.23 & 0.11 & 0.08 & 0.03 & 3    & 0.50 & 0.04 & 0.20 & 0.31 & 2   \\
         $\mathrm{[Y/Fe]}$  & 0.44 & 0.12 & 0.04 & 0.37 & 4    & 0.13 & 0.38 & 0.09 & 0.05 & 4    &-0.59 & 0.16 & 0.11 &-0.29 & 4   \\
         $\mathrm{[Zr/Fe]}$ & 0.80 & 0.11 & 0.08 & --   & 16   & 0.30 & 0.15 & 0.13 & --   & 16   &-0.15 & 0.11 & 0.12 & --   & 14  \\
         $\mathrm{[Nb/Fe]}$ & 1.01 & --   & 0.42 & --   & 1    & --   & --   & --   & --   & --   & 0.25 & --   & --   & --   & 1   \\
         $\mathrm{[Mo/Fe]}$ & 0.98 & --   & 0.09 & --   & 1    & 0.68 & --   & 0.10 & --   & 1    & --   & --   & --   & --   & --  \\
         $\mathrm{[Ru/Fe]}$ & 1.44 & 0.06 & 0.06 & --   & 2    & --   & --   & --   & --   & --   & --   & --   & --   & --   & --  \\
         $\mathrm{[Rh/Fe]}$ & 1.42 & 0.06 & 0.12 & --   & 2    & --   & --   & --   & --   & --   & --   & --   & --   & --   & --  \\
         $\mathrm{[Pd/Fe]}$ & 0.41 & 0.02 & 0.04 & --   & 4    & --   & --   & --   & --   & --   & 0.03 & 0.01 & 0.02 & --   & 4   \\
         $\mathrm{[Ba/Fe]}$ & 1.03 & 0.04 & 0.11 & 0.97 & 2    &-0.16 & 0.11 & 0.06 &-0.24 & 3    &-1.42 & 0.20 & 0.13 &-1.00 & 4   \\
         $\mathrm{[La/Fe]}$ & 1.41 & 0.09 & 0.06 & --   & 13   & 0.34 & 0.10 & 0.06 & --   & 14   &-0.79 & 0.19 & 0.17 & --   & 4   \\
         $\mathrm{[Ce/Fe]}$ & 1.11 & 0.08 & 0.06 & --   & 35   & 0.40 & 0.11 & 0.14 & --   & 25   &-0.80 & 0.16 & 0.18 & --   & 13  \\
         $\mathrm{[Pr/Fe]}$ & 1.44 & 0.06 & 0.06 & --   & 10   & 0.45 & 0.10 & 0.12 & --   & 8    &-0.59 & 0.12 & 0.22 & --   & 6   \\
         $\mathrm{[Nd/Fe]}$ & 1.43 & 0.07 & 0.06 & --   & 61   & 0.54 & 0.13 & 0.08 & --   & 44   &-0.64 & 0.04 & 0.19 & --   & 16  \\
         $\mathrm{[Sm/Fe]}$ & 1.60 & 0.09 & 0.07 & --   & 26   & 0.60 & 0.14 & 0.13 & --   & 13   &-0.33 & 0.10 & 0.19 & --   & 6   \\
         $\mathrm{[Eu/Fe]}$ & 1.79 & 0.06 & 0.05 & --   & 6    & 0.54 & 0.12 & 0.12 & --   & 3    &-0.73 & 0.05 & 0.16 & --  & 2   \\
         $\mathrm{[Gd/Fe]}$ & 1.62 & 0.10 & 0.05 & --   & 34   & 0.35 & 0.12 & 0.17 & --   & 14   &-0.53 & 0.09 & 0.20 & --   & 15  \\
         $\mathrm{[Tb/Fe]}$ & 1.63 & 0.09 & 0.04 & --   & 3    & 0.41 & 0.09 & 0.29 & --   & 2    & --   & --   & --   & --   & --  \\
         $\mathrm{[Dy/Fe]}$ & 1.83 & 0.07 & 0.07 & --   & 13   & 0.58 & 0.16 & 0.08 & --   & 14   &-1.01 & 0.07 & 0.09 & --   & 14  \\
         $\mathrm{[Ho/Fe]}$ & 1.63 & 0.12 & 0.06 & --   & 12   & 0.48 & 0.33 & 0.23 & --   & 8    &-1.05 & --   & 0.08 & --   & 1   \\
         $\mathrm{[Er/Fe]}$ & 1.82 & 0.20 & 0.07 & --   & 7    & 0.41 & 0.23 & 0.12 & --   & 6    &-0.34 & 0.11 & 0.21 & --   & 3   \\
         $\mathrm{[Tm/Fe]}$ & 1.75 & 0.09 & 0.07 & --   & 4    & 0.45 & --   & 0.09 & --   & 1    & --   & --   & --   & --   & --  \\
         $\mathrm{[Yb/Fe]}$ & 0.91 & --   & 0.13 & --   & 1    &-0.52 & --   & 0.06 & --   & 1    & --   & --   & --   & --   & --  \\
         $\mathrm{[Lu/Fe]}$ & 1.82 & --   & 0.04 & --   & 1    & --   & --   & --   & --   & --   &-0.60 & --   & 0.14 & --   & 1   \\
         $\mathrm{[Hf/Fe]}$ & 1.40 & 0.07 & 0.08 & --   & 5    & --   & --   & --   & --   & --   & --   & --   & --   & --   & --  \\ 
         $\mathrm{[Os/Fe]}$ & 1.87 & 0.11 & 0.14 & --   & 3    & --   & --   & --   & --   & --   & --   & --   & --   & --   & --  \\
         $\mathrm{[Ir/Fe]}$ & 1.49 & 0.01 & 0.08 & --   & 2    & 0.38 & 0.09 & 0.10 & --   & 2    & --   & --   & --   & --   & --  \\
         $\mathrm{[Pb/Fe]}$ & 0.51 & --   & 0.24 & --   & 1    & 0.41 & --   & 0.19 & --   & 1    &-0.97 & --   & --   & --   & 1   \\
         $\mathrm{[Th/Fe]}$ & 1.89 & 0.05 & 0.08 & --   & 4    & 1.20 & --   & --   & --   & 1    & --   & --   & --   & --   & --  \\
    \hline
    \end{tabular}
    \begin{tablenotes}
        \item \textit{Notes}. $\sigma_{stat}$ is the statistical uncertainty, only presented if more than one line is measured. $\sigma_{par}$ is the uncertainty varying the atmospheric parameters in their uncertainties. Upper limits are show as both the $\sigma_{stat}$ and $\sigma_{par}$ as --. Elements not possible to measure have the three columns as --. $\star$ With carbon correction, [C/Fe]=0.06~dex. $\dagger$Using non-LTE corrections from \citet{Nordlander2017Non-LTEStars}, the non-LTE abundances would be $-$0.23 for \BPSCS, $-$0.31 for \TYC, and $-$0.05 for \HD.
    \end{tablenotes}
    \end{threeparttable}    
\end{table*}
\begin{table*}[]
    \centering
    \begin{threeparttable}
    \caption{Kinematic and dynamic parameters}
    \label{tab:kinematics}
    \begin{tabular}{lcccccccc}
    \hline
    \hline
        Star & \En & \Lz & $J_x$ & $J_y$ & \zmax & $R_{apo}$ & $R_{peri}$ &  Eccentricity \\ 
         ~   & ($km^2/s^2$) & ($km/s/kpc$) & ~ & ~ & (kpc) & (kpc) & (kpc) & ~ \\
        \hline
        SMSS J175046.30-425506.9 & -92,863 & -1,368 & -0.26 & -0.34 & 37.11 & 49.67 & 5.62 & 0.80 \\ 
        BPS CS 29497-0004 & -140,517 & 1,150 & 0.54 & -0.20 & 8.01 & 17.20 & 3.75 & 0.64 \\ 
        BD-16 251 & -134,512 & -1,449 & -0.54 & 0.28 & 12.84 & 16.63 & 8.00 & 0.35 \\ 
        HE 1523-0901 & -142,700 & -1,494 & -0.67 & 0.02 & 8.23 & 15.02 & 5.63 & 0.45 \\ 
        UCAC4 392-119337 & -126,956 & -1,340 & -0.46 & 0.06 & 16.27 & 21.96 & 5.94 & 0.57 \\ 
        LP 584-32 & -138,362 & -1,095 & -0.41 & 0.08 & 15.70 & 20.19 & 6.03 & 0.47 \\ 
        HE 2134-3940 & -176,588 & -1,045 & -0.86 & 0.07 & 2.39 & 6.72 & 4.23 & 0.23 \\ 
        BPS CS 31062-0050 & -171,700 & 274 & 0.27 & -0.66 & 3.51 & 9.89 & 1.11 & 0.80 \\ 
        HE 1305+0007 & -115,939 & -275 & -0.10 & -0.70 & 17.47 & 30.67 & 1.96 & 0.88 \\ 
        CD-28 1082 & -168,959 & 987 & 0.74 & 0.02 & 3.75 & 8.68 & 3.66 & 0.41 \\ 
        2MASS J19280402-5924264 & -181,734 & 13 & 0.02 & -0.18 & 5.44 & 7.42 & 1.07 & 0.76 \\ 
        HD 209621 & -159,917 & -1,197 & -0.75 & -0.21 & 1.59 & 11.32 & 3.41 & 0.54 \\ 
        HE 1105+0027 & -161,383 & 1,145 & 0.75 & -0.07 & 3.46 & 10.54 & 3.71 & 0.48 \\ 
        HE 0338-3945 & -173,546 & 571 & 0.53 & -0.27 & 3.23 & 9.06 & 1.75 & 0.68 \\ 
        BPS CS 22898-0027 & -184,056 & 169 & 0.22 & -0.77 & 1.30 & 8.04 & 1.39 & 0.76 \\ 
        HE 2148-1247 & -140,343 & 69 & 0.03 & 0.59 & 15.57 & 16.06 & 5.22 & 0.51 \\ 
        CD-24 266 & -172,490 & -521 & -0.53 & -0.38 & 1.74 & 9.34 & 2.28 & 0.64 \\ 
        HE 0448-4806 & -175,811 & 729 & 0.65 & -0.32 & 0.82 & 8.77 & 2.02 & 0.63 \\ 
        BPS CS 29497-034 & -157,344 & -1,617 & -0.90 & 0.09 & 3.27 & 8.77 & 7.18 & 0.10 \\ 
        LP 625-44 & -149,376 & -846 & -0.00 & -1.00 & 6.81 & 26.68 & 3.18 & 0.68 \\ 
        HE 0058-0244 & -178,475 & -193 & -0.23 & -0.65 & 3.55 & 8.80 & 0.84 & 0.83 \\ 
        HE 2122-4707 & -177,420 & -660 & -0.58 & 0.41 & 4.53 & 5.94 & 4.48 & 0.14 \\ 
        HE 0010-3422b & -154,789 & -18 & -0.01 & 0.45 & 12.60 & 12.62 & 2.81 & 0.64 \\ 
        HE 0243-3044 & -162,394 & 394 & 0.33 & -0.15 & 7.27 & 11.19 & 1.71 & 0.74 \\ 
        CD-27 14351 & -191,072 & 250 & 0.35 & -0.42 & 2.97 & 6.73 & 0.88 & 0.77 \\ 
        HE 2339-0837 & -71,901 & 1,166 & 0.13 & -0.01 & 77.38 & 80.92 & 11.80 & 0.74 \\ 
        TYC 8100-833-1 & -160,751 & 36 & 0.03 & -0.88 & 4.08 & 12.33 & 0.91 & 0.86 \\ 
        SMSS J145341.38+004046.7 & -129,002 & 67 & 0.03 & 0.14 & 21.49 & 21.73 & 9.82 & 0.38 \\ 
        TYC 7325-920-1 & -120,416 & 2,109 & 0.63 & -0.14 & 12.47 & 25.39 & 6.57 & 0.59 \\ 
        TYC 8444-76-1 & -184,368 & 250 & 0.31 & -0.24 & 4.30 & 7.15 & 1.11 & 0.73 \\ 
        \hline
        BPS CS 29529-0089 & -166,577 & 1,100 & 0.78 & -0.04 & 2.99 & 9.31 & 3.70 & 0.43 \\ 
        TYC 9219-2422-1 & -133,612 & 812 & 0.36 & -0.64 & 0.78 & 21.10 & 2.19 & 0.81 \\ 
        HD 122563 & -172,750 & -137 & -0.15 & -0.84 & 3.51 & 9.88 & 0.77 & 0.86 \\ 
        \hline
    \end{tabular}
    \begin{tablenotes}
        \item \textit{Notes}. Typical uncertainties are less 1\%.
    \end{tablenotes}
    \end{threeparttable} 
\end{table*}

\section{Hyperfine structure and isotopic splitting of the \ion{Eu}{II} lines.}\label{ap:eu}

This appendix lists the oscillator strengths for each hyperfine component of the \ion{Eu}{ii} lines that were analyzed in this work, for the two isotopes of importance that were considered here. The isotopic fractions (47.81\% and 52.19\%, for A = 151 and A = 153, respectively) are taken into account as an additional multiplicative factor internally to the spectrum synthesis.

\begin{table}
\caption{\label{tab.eu2_3724}Hyperfine structure of the \ion{Eu}{II} line at 3724.9 \AA.}
\centering
\begin{tabular}{cccc}
\hline\hline
\multicolumn{2}{c}{ A = 151} & \multicolumn{2}{c}{ A = 153} \\
Wavelength ($\lambda$) & $\log{gf}$ & Wavelength ($\lambda$) & $\log{gf}$ \\
\hline
3724.835 & $-$1.823 & 3724.903 & $-$1.346 \\
3724.835 & $-$1.346 & 3724.903 & $-$1.823 \\
3724.852 & $-$1.627 & 3724.911 & $-$1.823 \\
3724.852 & $-$1.288 & 3724.911 & $-$1.288 \\
3724.853 & $-$1.823 & 3724.912 & $-$1.627 \\
3724.876 & $-$1.568 & 3724.922 & $-$1.568 \\
3724.877 & $-$1.158 & 3724.923 & $-$1.627 \\ 
3724.877 & $-$1.627 & 3724.923 & $-$1.158 \\
3724.906 & $-$1.605 & 3724.936 & $-$1.605 \\
3724.908 & $-$1.007 & 3724.937 & $-$1.007 \\
3724.909 & $-$1.568 & 3724.937 & $-$1.568 \\
3724.943 & $-$1.791 & 3724.952 & $-$1.791 \\
3724.946 & $-$0.856 & 3724.953 & $-$0.856 \\
3724.947 & $-$1.605 & 3724.954 & $-$1.605 \\
3724.990 & $-$0.712 & 3724.972 & $-$0.712 \\
3724.992 & $-$1.791 & 3724.974 & $-$1.791 \\
\hline
\end{tabular}
\end{table}
\begin{table}
\caption{\label{tab.eu2_3819}Hyperfine structure of the \ion{Eu}{II} line at 3819.6 \AA.}
\centering
\begin{tabular}{cccc}
\hline\hline
\multicolumn{2}{c}{ A = 151} & \multicolumn{2}{c}{ A = 153} \\
Wavelength ($\lambda$) & $\log{gf}$ & Wavelength ($\lambda$) & $\log{gf}$ \\
\hline
3819.578 & $-$0.619 & 3819.645 & $-$0.619 \\
3819.595 & $-$0.510 & 3819.653 & $-$1.288 \\
3819.597 & $-$1.288 & 3819.654 & $-$0.510 \\
3819.619 & $-$0.401 & 3819.665 & $-$2.506 \\
3819.621 & $-$1.098 & 3819.665 & $-$1.098 \\
3819.623 & $-$2.506 & 3819.666 & $-$0.401 \\
3819.649 & $-$0.296 & 3819.680 & $-$0.296 \\
3819.652 & $-$1.044 & 3819.680 & $-$2.360 \\
3819.655 & $-$2.360 & 3819.681 & $-$1.044 \\
3819.685 & $-$0.197 & 3819.695 & $-$0.197 \\
3819.690 & $-$1.086 & 3819.698 & $-$1.086 \\
3819.694 & $-$2.447 & 3819.699 & $-$2.447 \\
3819.728 & $-$0.104 & 3819.712 & $-$0.104 \\
3819.734 & $-$1.276 & 3819.717 & $-$1.276 \\
3819.739 & $-$2.775 & 3819.720 & $-$2.775 \\
\hline
\end{tabular}
\end{table}
\begin{table}
\caption{\label{tab.eu2_3907}Hyperfine structure of the \ion{Eu}{II} line at 3907.1 \AA.}
\centering
\begin{tabular}{cccc}
\hline\hline
\multicolumn{2}{c}{ A = 151} & \multicolumn{2}{c}{ A = 153} \\
Wavelength ($\lambda$) & $\log{gf}$ & Wavelength ($\lambda$) & $\log{gf}$ \\
\hline
3907.046 & $-$0.376 & 3907.095 & $-$0.376 \\
3907.080 & $-$0.544 & 3907.111 & $-$0.544 \\
3907.093 & $-$1.188 & 3907.116 & $-$1.188 \\
3907.108 & $-$0.744 & 3907.124 & $-$0.744 \\
3907.119 & $-$1.022 & 3907.128 & $-$1.022 \\
3907.130 & $-$0.996 & 3907.133 & $-$0.996 \\
3907.131 & $-$2.285 & 3907.133 & $-$2.285 \\
3907.138 & $-$1.012 & 3907.137 & $-$1.012 \\
3907.147 & $-$1.360 & 3907.140 & $-$1.360 \\
3907.148 & $-$1.920 & 3907.142 & $-$1.920 \\
3907.152 & $-$1.098 & 3907.142 & $-$1.098 \\
3907.159 & $-$1.774 & 3907.145 & $-$1.263 \\
3907.160 & $-$1.263 & 3907.146 & $-$1.774 \\
3907.164 & $-$1.807 & 3907.148 & $-$1.807 \\
\hline
\end{tabular}
\end{table}
\begin{table}
\caption{\label{tab.eu2_3930}Hyperfine structure of the \ion{Eu}{II} line at 3930.5 \AA.}
\centering
\begin{tabular}{cccc}
\hline\hline
\multicolumn{2}{c}{ A = 151} & \multicolumn{2}{c}{ A = 153} \\
Wavelength ($\lambda$) & $\log{gf}$ & Wavelength ($\lambda$) & $\log{gf}$ \\
\hline
3930.425 & $-$1.222 & 3930.480 & $-$1.222 \\
3930.430 & $-$0.329 & 3930.485 & $-$0.329 \\
3930.470 & $-$1.047 & 3930.500 & $-$1.047 \\
3930.473 & $-$0.537 & 3930.501 & $-$0.537 \\
3930.478 & $-$1.222 & 3930.506 & $-$1.222 \\
3930.507 & $-$1.025 & 3930.517 & $-$0.774 \\
3930.508 & $-$0.774 & 3930.518 & $-$1.025 \\
3930.511 & $-$1.047 & 3930.518 & $-$1.047 \\
3930.536 & $-$1.101 & 3930.531 & $-$1.025 \\
3930.537 & $-$1.043 & 3930.531 & $-$1.043 \\
3930.539 & $-$1.025 & 3930.533 & $-$1.101 \\
3930.557 & $-$1.310 & 3930.541 & $-$1.101 \\
3930.558 & $-$1.319 & 3930.542 & $-$1.319 \\
3930.558 & $-$1.101 & 3930.544 & $-$1.310 \\
3930.570 & $-$1.407 & 3930.548 & $-$1.310 \\
3930.571 & $-$1.310 & 3930.550 & $-$1.407 \\
\hline
\end{tabular}
\end{table}
\begin{table}
\caption{\label{tab.eu2_4129}Hyperfine structure of the \ion{Eu}{II} line at 4129.7 \AA.}
\centering
\begin{tabular}{cccc}
\hline\hline
\multicolumn{2}{c}{ A = 151} & \multicolumn{2}{c}{ A = 153} \\
Wavelength ($\lambda$) & $\log{gf}$ & Wavelength ($\lambda$) & $\log{gf}$ \\
\hline
4129.615 & $-$1.512 & 4129.696 & $-$1.512 \\
4129.618 & $-$1.035 & 4129.698 & $-$1.035 \\
4129.632 & $-$1.316 & 4129.702 & $-$1.316 \\
4129.637 & $-$0.977 & 4129.705 & $-$0.977 \\
4129.640 & $-$1.512 & 4129.708 & $-$1.512 \\
4129.657 & $-$1.257 & 4129.712 & $-$1.257 \\
4129.663 & $-$0.847 & 4129.716 & $-$0.847 \\
4129.667 & $-$1.316 & 4129.719 & $-$1.316 \\
4129.690 & $-$1.294 & 4129.727 & $-$1.294 \\
4129.696 & $-$0.696 & 4129.730 & $-$0.696 \\
4129.702 & $-$1.257 & 4129.733 & $-$1.257 \\
4129.731 & $-$1.480 & 4129.748 & $-$1.480 \\
4129.738 & $-$0.545 & 4129.749 & $-$0.545 \\
4129.744 & $-$1.294 & 4129.751 & $-$1.294 \\
4129.788 & $-$0.401 & 4129.773 & $-$0.401 \\
4129.795 & $-$1.480 & 4129.774 & $-$1.480 \\
\hline
\end{tabular}
\end{table}
\begin{table}
\caption{\label{tab.eu2_4205}Hyperfine structure of the \ion{Eu}{II} line at 4205.0 \AA.}
\centering
\begin{tabular}{cccc}
\hline\hline
\multicolumn{2}{c}{ A = 151} & \multicolumn{2}{c}{ A = 153} \\
Wavelength ($\lambda$) & $\log{gf}$ & Wavelength ($\lambda$) & $\log{gf}$ \\
\hline
4204.895 & $-$1.112 & 4204.995 & $-$1.112 \\
4204.898 & $-$1.413 & 4204.996 & $-$1.413 \\
4204.903 & $-$2.368 & 4204.999 & $-$2.368 \\
4204.921 & $-$0.936 & 4205.006 & $-$0.936 \\
4204.926 & $-$1.230 & 4205.009 & $-$1.230 \\
4204.933 & $-$2.258 & 4205.013 & $-$2.258 \\
4204.958 & $-$0.773 & 4205.023 & $-$0.773 \\
4204.965 & $-$1.171 & 4205.027 & $-$1.171 \\
4204.974 & $-$2.368 & 4205.031 & $-$2.367 \\
4205.006 & $-$0.627 & 4205.045 & $-$0.627 \\
4205.015 & $-$1.205 & 4205.049 & $-$1.205 \\
4205.025 & $-$2.710 & 4205.053 & $-$2.710 \\
4205.065 & $-$0.496 & 4205.071 & $-$0.496 \\
4205.075 & $-$1.388 & 4205.075 & $-$1.388 \\
4205.134 & $-$0.376 & 4205.101 & $-$0.376 \\
\hline
\end{tabular}
\end{table}
\begin{table}
\caption{\label{tab.eu2_6645}Hyperfine structure of the \ion{Eu}{II} line at 6645.1 \AA.}
\centering
\begin{tabular}{cccc}
\hline\hline
\multicolumn{2}{c}{ A = 151} & \multicolumn{2}{c}{ A = 153} \\
Wavelength ($\lambda$) & $\log{gf}$ & Wavelength ($\lambda$) & $\log{gf}$ \\
\hline
6645.072 & $-$0.522 & 6645.074 & $-$1.828 \\
6645.079 & $-$1.828 & 6645.076 & $-$0.522 \\
6645.087 & $-$3.472 & 6645.076 & $-$3.472 \\
6645.098 & $-$0.598 & 6645.089 & $-$0.598 \\
6645.106 & $-$1.633 & 6645.091 & $-$1.633 \\
6645.113 & $-$3.154 & 6645.096 & $-$3.154 \\
6645.120 & $-$0.677 & 6645.099 & $-$0.677 \\
6645.128 & $-$1.588 & 6645.103 & $-$1.588 \\
6645.135 & $-$3.082 & 6645.106 & $-$0.760 \\
6645.138 & $-$0.760 & 6645.109 & $-$3.082 \\
6645.146 & $-$1.640 & 6645.111 & $-$0.844 \\
6645.152 & $-$3.250 & 6645.112 & $-$1.640 \\
6645.153 & $-$0.844 & 6645.116 & $-$0.926 \\
6645.160 & $-$1.835 & 6645.118 & $-$1.835 \\
6645.165 & $-$0.926 & 6645.118 & $-$3.250 \\
\hline
\end{tabular}
\end{table}
\section{Hyperfine structure and isotopic splitting  of the \ion{Ba}{II} lines.}\label{ap:ba}

This appendix lists the oscillator strengths for the isotopic splitting of the \ion{Ba}{ii} lines, and their hyperfine components when relevant. The isotopic fractions (2.417\%, 6.592\%, 7.854\%, 11.232\%, and 71.698\%, for A = 134, 135, 136, 137, and 138, respectively) are taken into account as an additional multiplicative factor internally to the spectrum synthesis.

\begin{table}
\caption{\label{tab.ba2_5853}Isotopic splitting and hyperfine structure of the \ion{Ba}{II} line at 5853.675 \AA.}
\centering
\begin{tabular}{ccc}
\hline\hline
Wavelength ($\lambda$) & Mass number & $\log{gf}$ \\
\hline
5853.677 & 134 & $-$0.907 \\
5853.676 & 135 & $-$0.907 \\
5853.675 & 136 & $-$0.907 \\
5853.669 & 137 & $-$1.965 \\
5853.671 & 137 & $-$2.111  \\
5853.671 & 137 & $-$1.907  \\
5853.673 & 137 & $-$2.111  \\
5853.673 & 137 & $-$2.509  \\
5853.675 & 137 & $-$1.810  \\
5853.676 & 137 & $-$1.907 \\ 
5853.676 & 137 & $-$1.363  \\
5853.674 & 138 & $-$0.907 \\
\hline
\end{tabular}
\end{table}

\begin{table}
\caption{\label{tab.ba2_6141}Isotopic splitting and hyperfine structure of the \ion{Ba}{II} line at 6141.71 \AA.}
\centering
\begin{tabular}{ccc}
\hline\hline
Wavelength ($\lambda$) & Mass number & $\log{gf}$ \\
\hline
6141.715 & 134 & $-$0.032 \\
6141.714 & 135 & $-$0.032 \\
6141.714 & 136 & $-$0.032 \\
6141.709 & 137 & $-$1.266  \\
6141.709 & 137 & $-$0.458 \\
6141.710 & 137 & $-$2.412 \\
6141.715 & 137 & $-$0.664 \\
6141.716 & 137 & $-$1.169 \\
6141.717 & 137 & $-$2.236 \\
6141.718 & 137 & $-$0.914 \\
6141.719 & 137 & $-$1.236 \\
6141.719 & 137 & $-$1.282 \\
6141.713 & 138 & $-$0.032 \\
\hline
\end{tabular}
\end{table}

\begin{table}
\caption{\label{tab.ba2_6496}Isotopic splitting and hyperfine structure of the \ion{Ba}{II} line at 6496.69 \AA.}
\centering
\begin{tabular}{ccc}
\hline\hline
Wavelength ($\lambda$) & Mass number & $\log{gf}$ \\
\hline
6496.900 & 134 & $-$0.407 \\
6496.899 & 135 & $-$1.912  \\
6496.899 & 135 & $-$1.213  \\
6496.899 & 135 & $-$0.766  \\
6496.899 & 135 & $-$1.611  \\
6496.899 & 135 & $-$1.213  \\
6496.899 & 135 & $-$1.213  \\
6496.900 & 136 & $-$0.407 \\
6496.883 & 137 & $-$1.912  \\
6496.888 & 137 & $-$1.213 \\ 
6496.896 & 137 & $-$0.766  \\
6496.902 & 137 & $-$1.611  \\
6496.904 & 137 & $-$1.213  \\
6496.909 & 137 & $-$1.213  \\
6496.898 & 138 & $-$0.407 \\
\hline
\end{tabular}
\end{table}

\section{Line fits}\label{ap:plots}
\begin{figure}[t]
    \centering
    \includegraphics[width=\linewidth]{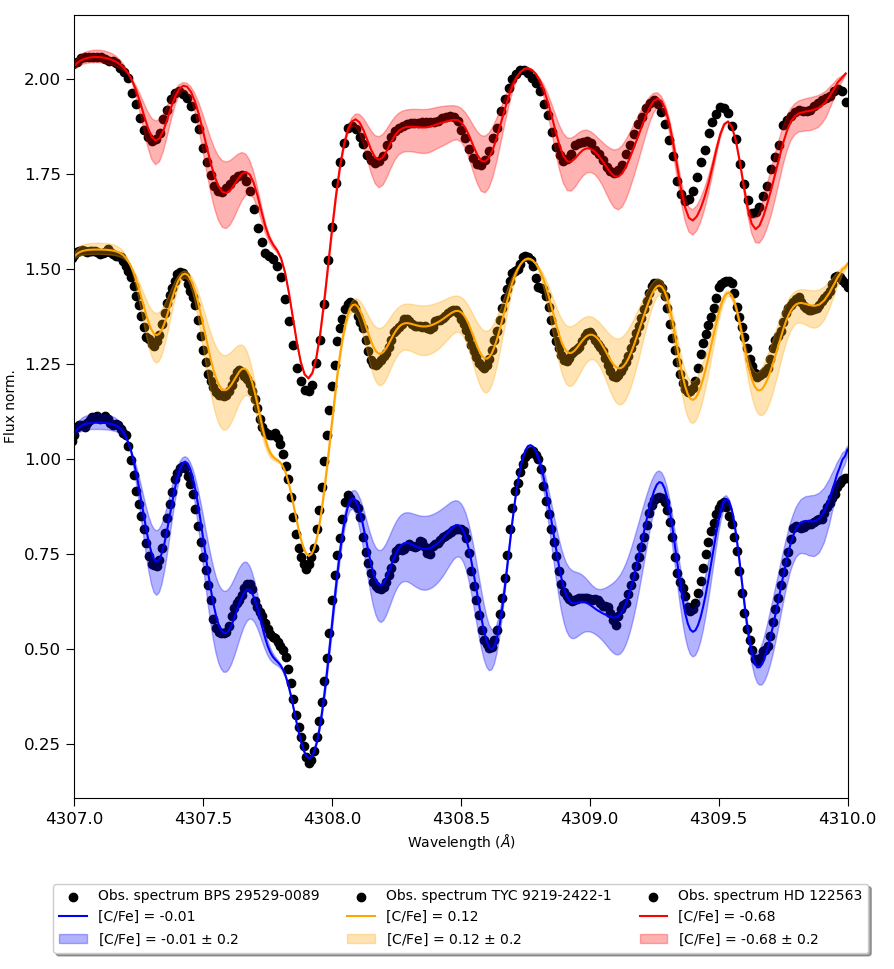}
    \caption{The molecular band of CH from 4307 to 4310 \AA~ for \BPSCS, \TYC~ and \HD~ with fitted abundance and $\pm$0.2 dex.}
    \label{fig:carbonband}
\end{figure}

\begin{figure}
    \centering
    \includegraphics[width=\linewidth, keepaspectratio]{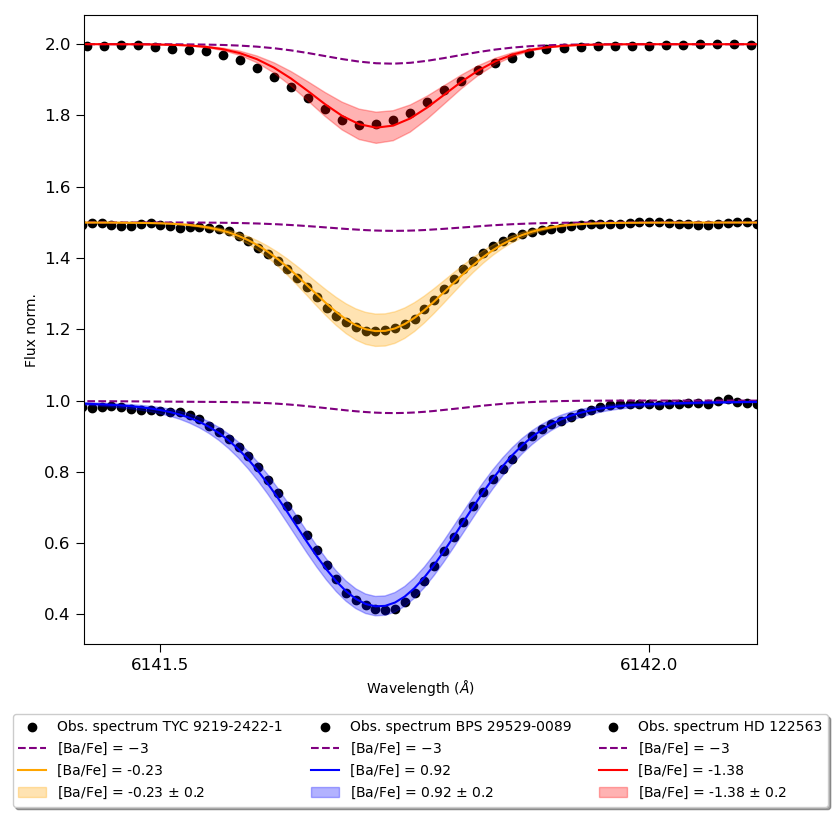}
    \caption{Ba line in 6141 \AA~ for \BPSCS, \TYC~ and \HD~ with fitted abundance and $\pm$0.2 dex. Fluxes are offset by 0.5. \BPSCS~ is shown in blue, \TYC~ in orange and \HD~ in red.}
    \label{fig:Ba}
\end{figure}

\begin{figure}
    \centering
    \includegraphics[width=\linewidth, keepaspectratio]{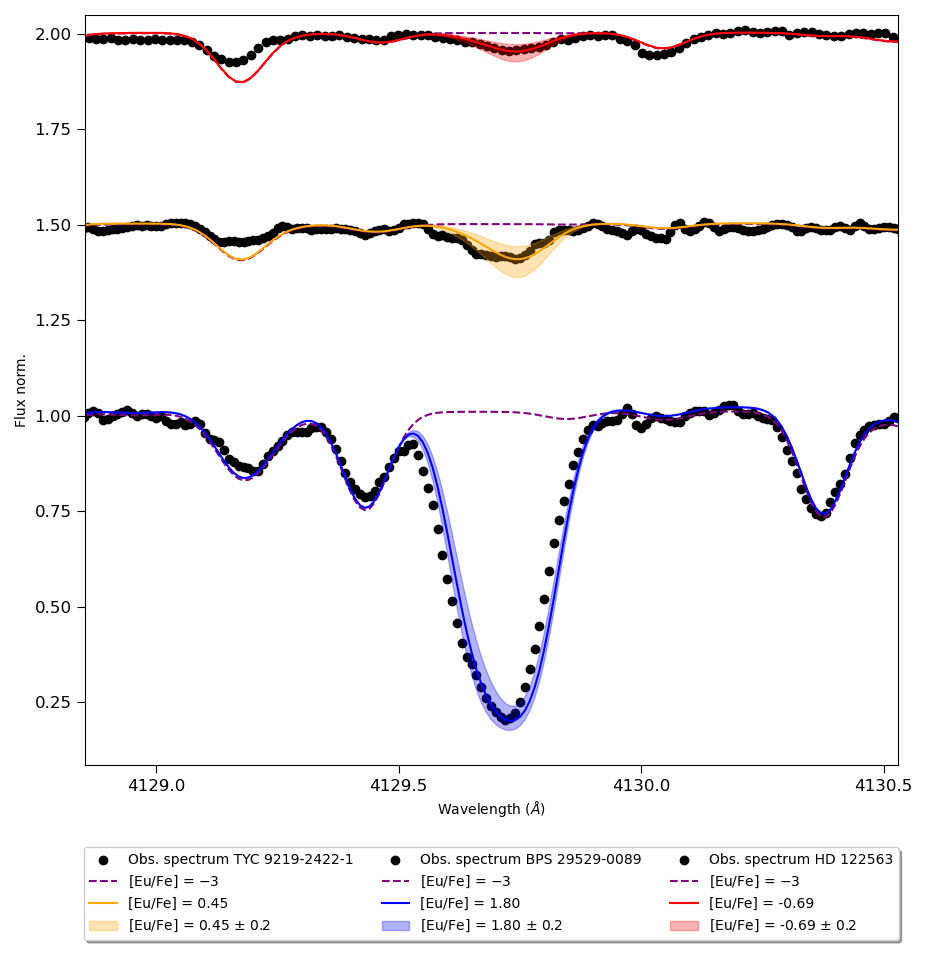}
    \caption{Eu line in 4129 \AA~ for \BPSCS, \TYC~ and \HD~ with fitted abundance and $\pm$0.2 dex. Fluxes are offset by 0.5. \BPSCS~ is shown in blue, \TYC~ in orange and \HD~ in red.}
    \label{fig:Eu}
\end{figure}

\begin{figure}
    \centering
    \includegraphics[width=\linewidth]{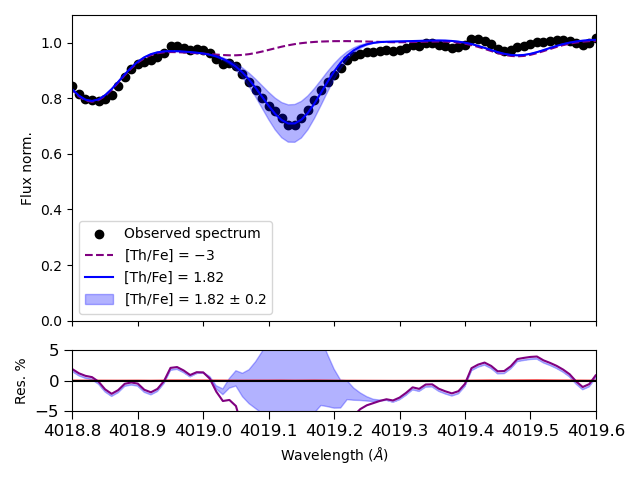}
    \caption{Th line 4019 \AA~ for \BPSCS~ with best fit abundance for the line and $\pm$0.2 dex.}
    \label{fig:BPS_Th}
\end{figure}

\begin{figure}
    \centering
    \includegraphics[width=\linewidth]{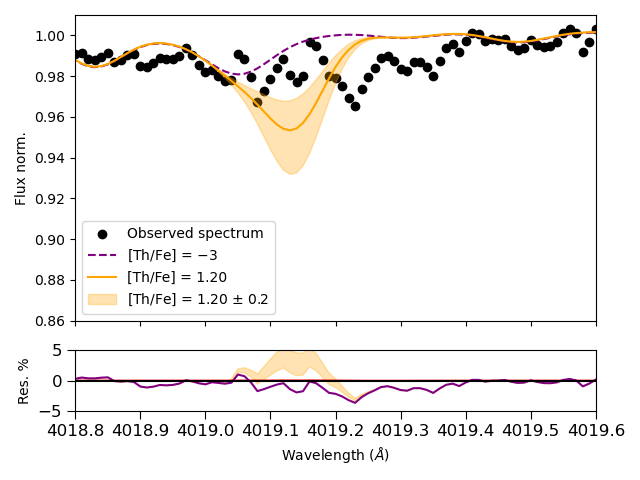}
    \caption{Th line 4019 \AA~ for \TYC~ with upper limit and $\pm$0.2 dex.}
    \label{fig:TYC_Th}
\end{figure}

\begin{figure}
    \centering
    \includegraphics[width=\linewidth]{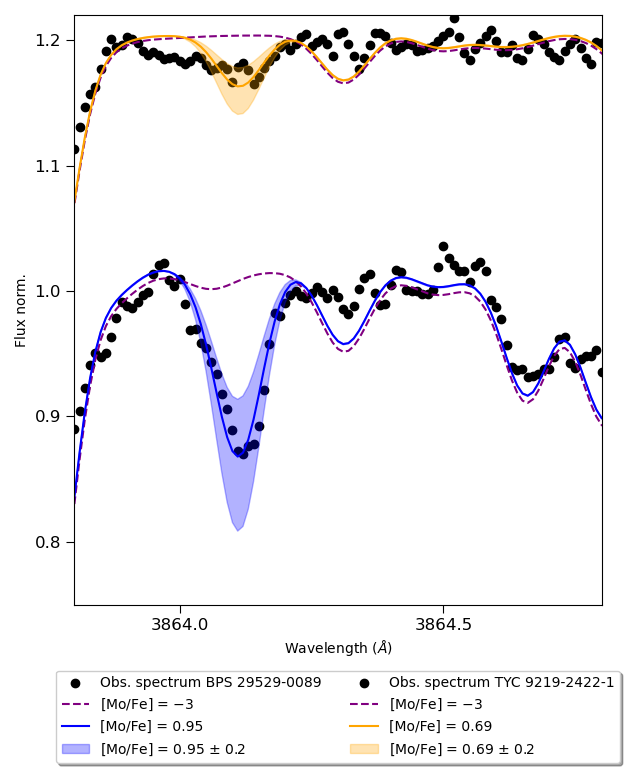}
    \caption{Mo line 3864 \AA~ for \BPSCS~ and \TYC~ with fitted abundance and $\pm$0.2 dex. Fluxes are offset by 0.2.}
    \label{fig:Mo}
\end{figure}

\begin{figure}
    \centering
    \includegraphics[width=\linewidth]{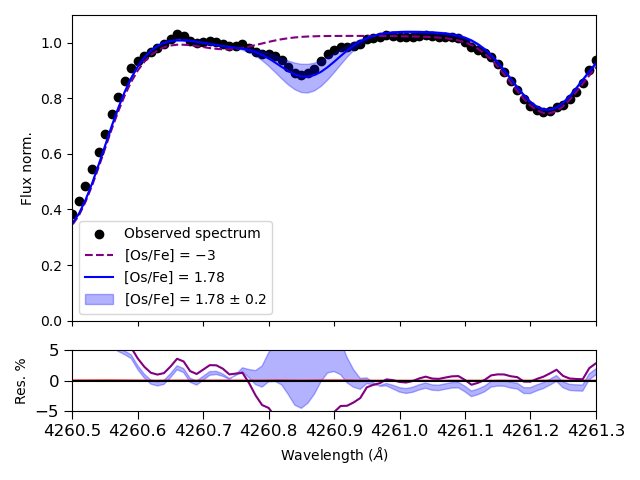}
    \caption{Os line in 4260 \AA~ for \BPSCS~ with fitted abundance and $\pm$0.2 dex.}
    \label{fig:BPS_Os}
\end{figure}

\begin{figure}
    \centering
    \includegraphics[width=\linewidth]{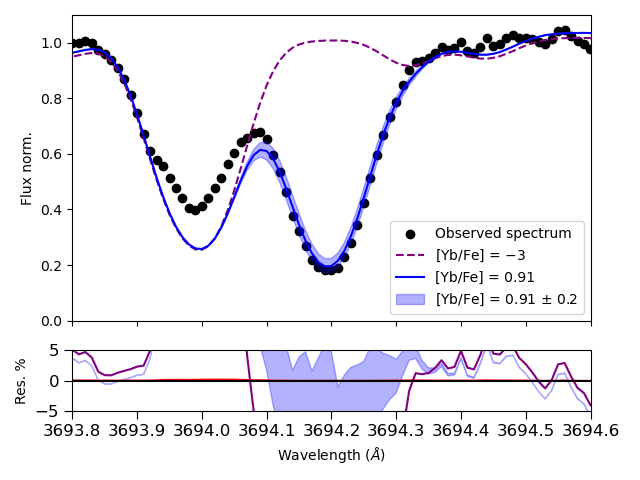}
    \caption{Yb line 3694 \AA~ for \BPSCS~ with fitted abundance and $\pm$0.2 dex.}
    \label{fig:BPS_Yb}
\end{figure}

\begin{figure}
    \centering
    \includegraphics[width=\linewidth]{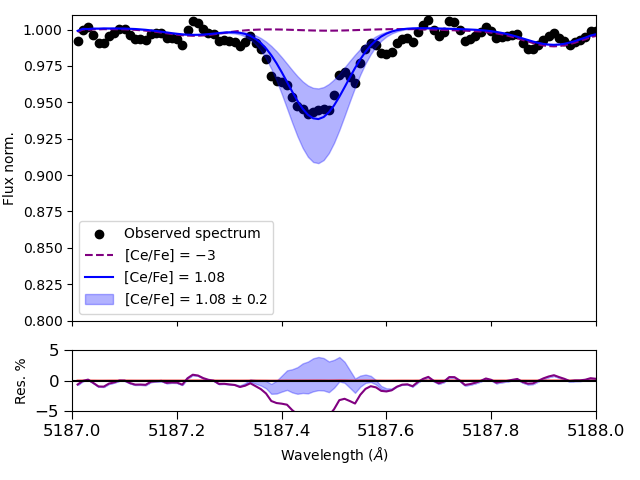}
    \caption{Ce line 5187 \AA~ for \BPSCS~ with fitted abundance and $\pm$0.2 dex.}
    \label{fig:BPS_Ce}
\end{figure}

\begin{figure}
    \centering
    \includegraphics[width=\linewidth]{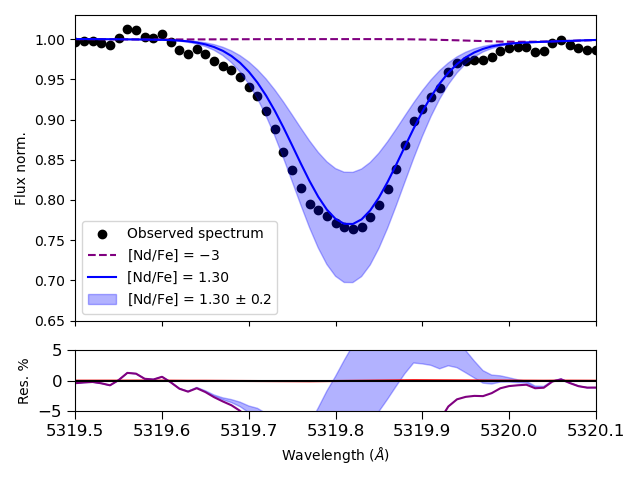}
    \caption{Nd line 5319 \AA~ for \BPSCS~ with fitted abundance and $\pm$0.2 dex.}
    \label{fig:BPS_Nd}
\end{figure}

\begin{figure}
    \centering
    \includegraphics[width=\linewidth]{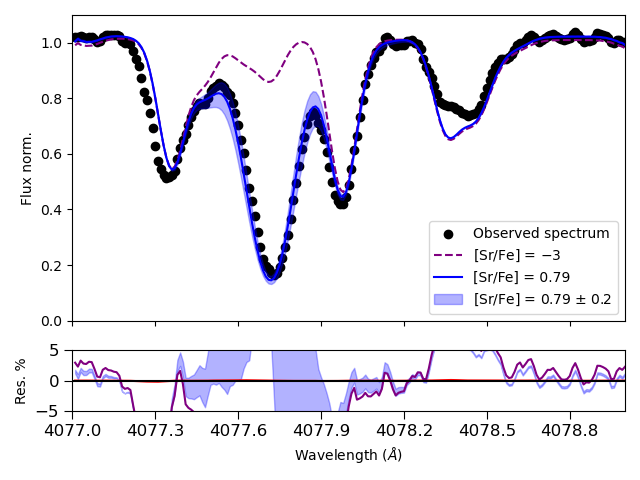}
    \caption{Sr line 4078 \AA~ for \BPSCS~ with fitted abundance and $\pm$0.2 dex.}
    \label{fig:BPS_Sr}
\end{figure}

On this appendix, we provide some line fits using the observed spectra and the synthetic fit with determined abundances. We provide the the fits for the carbon-band from 429 to 431 nm in Fig.~\ref{fig:carbonband}. 
Barium lines in Fig. \ref{fig:Ba}.
Europium lines in Fig. \ref{fig:Eu}.
Thorium lines in Figs. \ref{fig:BPS_Th} and \ref{fig:TYC_Th}.
Molybdenum in Fig. \ref{fig:Mo}.
For \BPSCS~ some other interesting element lines in Figs. Os: \ref{fig:BPS_Os}, Yb: \ref{fig:BPS_Yb}, Ce: \ref{fig:BPS_Ce}, Nd: \ref{fig:BPS_Nd} and Sr: \ref{fig:BPS_Sr}.

\end{appendix}
\end{document}